\documentclass[9pt,twocolumn,twoside]{optica}
\setboolean{shortarticle}{false}
\setboolean{minireview}{false}
\setboolean{displaycopyright}{false}

\usepackage{siunitx}

\title{High resolution and sensitivity bi-directional x-ray phase contrast imaging using 2D Talbot array illuminators}

\author[1,a,*]{Alex Gustschin}
\author[1,2,*]{Mirko Riedel}
\author[1]{Kirsten Taphorn}
\author[1]{Christian Petrich}
\author[1]{Wolfgang Gottwald}
\author[1]{Wolfgang Noichl}
\author[1]{Madleen Busse}
\author[3]{Sheila E. Francis}
\author[2]{Felix Beckmann}
\author[2]{Jörg U. Hammel}
\author[2]{Julian Moosmann}
\author[4]{Pierre Thibault}
\author[1]{Julia Herzen}

\affil[1]{Department of Physics and Munich School of Bioengineering, Technical University of Munich, 85748, Garching, Germany.}
\affil[2]{Institute of Materials Physics, Helmholtz-Zentrum hereon, Max-Planck-Str. 1, 21502 Geesthacht, Germany.}
\affil[3]{Department of Infection, Immunity and Cardiovascular Disease, Medical School, University of Sheffield S10 2RX, UK}
\affil[4]{Department of Physics, University of Trieste, Trieste 34217, Italy}

\affil[a]{Corresponding author: alex.gustschin@ph.tum.de}
\affil[*]{These authors contributed equally to this work}

\begin{abstract}
Two-dimensional Talbot array illuminators (TAIs) were designed, fabricated, and evaluated for high-resolution high-contrast x-ray phase imaging of soft tissue at 10-20\,keV. The TAIs create intensity modulations with a high compression ratio on the micrometer scale at short propagation distances. Their performance was compared with various other wavefront markers in terms of period, visibility, flux efficiency and flexibility to be adapted for limited beam coherence and detector resolution. Differential x-ray phase contrast and dark-field imaging were demonstrated with a one-dimensional, linear phase stepping approach yielding two-dimensional phase sensitivity using Unified Modulated Pattern Analysis (UMPA) for phase retrieval. The method was employed for x-ray phase computed tomography reaching a resolution of 3\,µm on an unstained murine artery. It opens new possibilities for three-dimensional, non-destructive, and quantitative imaging of soft matter such as virtual histology. The phase modulators can also be used for various other x-ray applications such as dynamic phase imaging, super-resolution structured illumination microscopy, or wavefront sensing.   
\end{abstract}

\begin{document}
\maketitle
\section{Introduction}
Various imaging techniques based on x-rays have opened unique insights into three-dimensional structures at the micro- and nanometer scale and even enabled the capture of time-resolved volumetric data due to recent innovations in x-ray sources, optics, detectors, high precision metrology, and advanced post-processing and reconstruction algorithms. Phase contrast techniques have become indispensable due to their capability to generate superior contrast in soft tissue compared to conventional attenuation-based mechanisms \cite{Fitzgerald2000}. While propagation-based methods provide good edge visibility \cite{Gureyev2009}, analyzer-based \cite{Davis1995}, interferometric \cite{Momose2003e, Pfeiffer2006e}, aperture-based \cite{Olivo2007b, Morgan2011} and speckle-based methods \cite{Berujon2012, Morgan2012, Zanette2014} are more sensitive to phase gradients and can deliver quantitative information for the separation of phase and absorption interaction. The latter techniques rely on various diffractive and absorptive beam modulator optics creating a defined intensity pattern after propagation in space. This modulation is altered by absorption, refraction, and scattering by the investigated object in the beam path. Various techniques have been successfully implemented to retrieve those different interactions from a sample and a reference scan both in single-shot mode \cite{Wen2010, Morgan2012, Morgan2013, Zanette2014} and from multiple exposures with stepped modulators \cite{Berujon2012, Berujon2017a, Zdora2017a, Zdora2018}. In speckle-based Imaging (SBI), a random modulation is introduced by a diffuser (e.g. sandpaper with a fine grain size), while other techniques generate periodic modulations with gratings or other diffractive or refractive arrays. In order to perform efficiently, the modulators have to generate a pattern with good contrast (visibility) and average feature sizes resolvable by the detector. 

In general, a stable and high-resolution bi-directional phase retrieval requires every detector pixel to undergo a high contrast modulation in both directions during phase stepping. For random speckle patterns, this requires a large number of stepped frames at the cost of longer acquisition times, higher radiation dose, and complexity in data handling and image processing \cite{Zdora2018}. Using a regular beam modulator (e.g. a grating pattern or a refractive lens array) and applying an adapted sampling scheme can avoid these problems and reduce the number of frames required to reach a high resolution and sensitivity.  

Currently, a remarkable effort is being undertaken to create such periodic diffractive optical elements (DOE) for a variety of x-ray applications. Reich et al. \cite{Reich2018} created an array of stacked compound refractive lenses (CRL) with a period of \SI{65}{\micro\meter}. Dos Santos Rolo et al. \cite{Rolo2018} fabricated a Shack-Hartman array with $20 \times 20$ micro-lenslets by 3D direct laser writing with a periodicity of \SI{50}{\micro\meter}. Kagias et al. \cite{Kagias2019} fabricated circular phase arrays for omnidirectional dark-field imaging with a unit cell period of \SI{80}{\micro\meter}. Mamyrbayev et al. \cite{Mamyrbayev2020} developed a 2D CRL array for sub-pixel resolution scanning transmission microscopy with a period of \SI{55}{\micro\meter}. As some of these recent examples show, the periods of such x-ray optics are still in the range of several tens of micrometers limiting the achievable performance in sensitivity and resolution. Different types of 2D gratings \cite{Morgan2013, Morgan2011, Zakharova2018, Sato2011, Itoh2011, Rutishauser2013b, Zanette2010} have been used with significantly smaller periods. However, they did not achieve comparable visibilities and flux efficiencies as the aforementioned beam modulators, which create periodic sharp foci in the detection plane. 

Considering multiple factors related to grating fabrication and instrumental limitations, we propose and demonstrate a 2D periodic phase-shifting grating for the x-ray regime, also known as Talbot array illuminator (TAI) from visible light literature \cite{Suleski1997, Arrizon1994, Szwaykowski1993}. Compared to previously described methods employing 2D phase gratings \cite{Morgan2013, Sato2011, Itoh2011, Morimoto2015} we have adapted a design that creates periodic foci with a higher compression ratio compared to e.g. checkerboard 2D modulators \cite{Morgan2013} or orthogonally stacked 1D linear gratings \cite{Morimoto2015}. In contrast to absorptive 2D gratings or Hartmann masks previously demonstrated for x-rays \cite{Morgan2011, Zakharova2019a, Rix2019}, the proposed TAI uses the entire transmitting radiation to generate the desired modulation. Compared to state-of-the-art refractive micro-lens arrays \cite{Reich2018, Rolo2018, Mikhaylov2020}, the fabricated phase arrays have a much larger field-of-view (FoV), are resistant to long and high radiation dose exposures and can be easily fabricated with up to one order of magnitude smaller periods (e.g. \SI{5}{\micro\meter}). Unlike random phase modulators (diffusers) used in SBI, the TAI can be tailored for optimal performance at a certain source coherence, spectral range, propagation distance, and detector point spread function (PSF). All of these aspects become crucial when the method is translated from coherent sources at large synchrotron facilities to laboratory-based micro-focus sources with polychromatic spectra.

In the present research, we evaluate customary designed TAIs of different periods and compare their visibility performance with a sandpaper diffuser at different propagation distances. Further, a 1D stepping scheme yielding bi-directional sensitivity is employed and compared with the random modulator for different numbers of phase steps. High-resolution bi-directional phase and dark-field imaging are demonstrated and a computed tomography (CT) phase scan of a murine artery embedded in paraffin is acquired. The proposed TAIs and acquisition schemes facilitate current state-of-the-art x-ray phase tomography, providing a convenient pathway for non-destructive, quantitative high-resolution 3D virtual histology.       

\begin{figure}[h!]
\centering
\includegraphics[width=\linewidth]{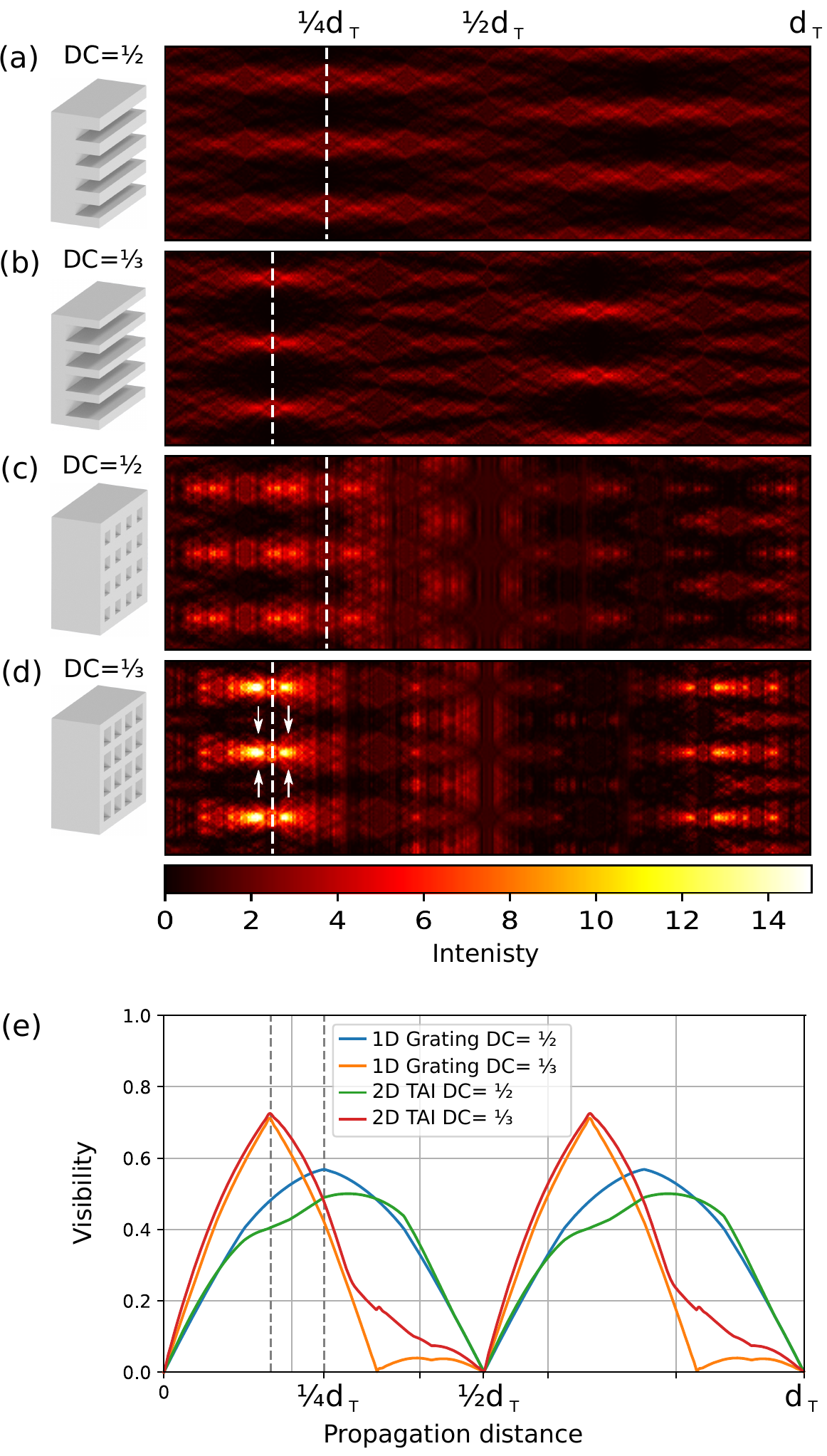}
\caption{Simulated Talbot carpets for (a) 1D linear grating with symmetric duty cycle (DC) and phase shift $\varphi=\pi/2$, (b) 1D linear grating with $DC =1/3$ and $\varphi=2\pi/3$, (c) 2D TAI with $DC=1/2$ and $\varphi=\pi/2$, (d) 2D TAI with $DC=1/3$ and $\varphi=2\pi/3$. The intensity is normalized to 1 in the grating plane. (e) Visibility with increasing propagation distance for the gratings (a-d). The fractional Talbot distances at which the first binary intensity modulation occur are denoted by dashed lines. Note that the second intensity modulation of the 2D TAIs is not visible in the Talbot carpets as they are shifted out of the plotted plane by half period.}  
\label{fig:Fig1}
\end{figure}

\section{Design of Talbot Array Illuminators}
\label{sec:design}

Current high-resolution x-ray detectors are thin scintillator screens focused with magnifying optics and coupled to CCD or CMOS pixel sensors, providing effective pixel sizes below \SI{1}{\micro\meter} and a spatial resolution in the range of \numrange{1}{2} \si{\micro\meter}. The task of creating the most efficient modulator consists of finding an optimal trade-off between the smallest possible period and the highest intensity contrast achievable with the PSF of the used detector. At the same time, the optics should attenuate the beam as little as possible which makes phase arrays fabricated from thin, x-ray transparent materials such as silicon the first choice. A broad variety of such periodic phase modulators has been studied \cite{Szwaykowski1993, Arrizon1994, Suleski1997} analytically to predict binary modulations at certain fractions of the Talbot distance $d_T = 2p^2/\lambda$, where $p$ is the period of the array and $\lambda$ the wavelength of the radiation. The highest theoretically achievable binary modulation with binary (two height levels) 1D linear phase gratings has a compression ratio of 1:3 \cite{Suleski1997}, i.e. the entire radiation is focused onto lines with a width of $1/3p$. A high compression ratio directly results in high visibility defined by: 
\begin{equation}
V = \frac{I_{\mathrm{max}}-I_{\mathrm{min}}}{I_{\mathrm{max}}+I_{\mathrm{min}}},
\label{eq:visibility}
\end{equation}

where $I_{\mathrm{max}}$ and $I_{\mathrm{min}}$ denote the maximal and minimal intensity within one modulation period. The measured visibility will be reduced by the detector blur which can be modeled by a convolution of the propagated intensity distribution with the PSF of the detector. 
In order to compare the performance of different grating designs, we calculated the resulting intensity patterns using the Fresnel-Kirchhoff diffraction formula (see Supplement 1 for details). The so-called Talbot carpets plotted in Figure \ref{fig:Fig1} (a-d) show how the spatial intensity evolves with the propagation distance. For binary gratings, it depends mainly on the duty cycle $DC$ (ratio of the phase-shifting fraction of the period) and the phase shift $\varphi$ of the grating profile. In most literature employing 2D phase gratings with x-rays (e.g. \cite{Morgan2013, Sato2011, Morimoto2015, Itoh2011}) symmetric duty cycles ($DC=0.5$) were used which also result in symmetric intensity modulation at fractional Talbot distances. However, a much stronger contrast can be achieved with asymmetric DC configurations when a convenient phase shift is chosen. Figure 1 (a) and (b) show calculated Talbot carpets of 1D linear phase gratings illustrating this intensity focusing effect. While the grating with symmetric duty cycle (a) produces an intensity modulation with a compression ratio of 1:2 at $1/4d_T$, the grating with $DC = 1/3$ and $\varphi= 2/3\pi$ (b) shows a stronger focusing with a compression ratio of 1:3 at an even shorter propagation distance ($1/6 d_T$). There are also several other asymmetric grating parameter configurations that create the same effect at different propagation distances \cite{Suleski1997}.

A wave propagation with two-dimensional arrays shows that this principle can be directly extended to a respective two-level 2D modulator creating a binary modulation with a compression ratio of 1:9. In Figure \ref{fig:Fig1}(c) and (d) the Talbot carpets for the 2D TAIs with the respective duty cycle designs are shown. As visible from the intensity (note that (a-d) have the same intensity color map), the 2D modulators result in an overall stronger focusing, and therefore, higher contrast is achieved compared to the 1D gratings. Further, the 2D TAI with $DC=1/3$ (d) shows a significantly higher intensity than its symmetric counterpart (c). Besides that, even stronger modulations (depicted by arrows) before and after the binary modulation at $1/6d_T$ are observable.

To quantify the visibility gain using the asymmetric gratings compared to the conventional ones, a plot of the visibility with propagation distance is provided in Figure 1(e). The visibility values have been calculated by equation \ref{eq:visibility} after convolving the intensity pattern at each propagation distance with a Gaussian 2D Kernel of $\sigma = 0.2p$ (accounting for PSF). The plot shows that both 1D and 2D modulators perform similarly in terms of visibility, although the 2D TAIs reach a higher compression ratio. That is comprehensible, as 2D- focused spots are affected stronger by the PSF blur compared to 1D linear intensity distributions. However, the advantage of the asymmetric modulators, both in 1D and 2D cases, is visible. The asymmetric 2D TAI reaches about a 40\% higher visibility than its symmetric counterpart at respective peak performance.

It is noteworthy that even stronger compression ratios can be achieved with binary phase arrays using $DC<1/3$. Although the created intensity pattern will not be binary, most of the intensity will be still focused on very narrow points \cite{Arrizon1994}. More complicated phase modulators e.g. with more than two height levels and sub-periodic features \cite{Szwaykowski1993} or other non-binary \cite{Yaroshenko2014c} (e.g. triangular, trapezoidal, or sinusoidal) DOE can also create stronger focusing than conventional binary phase gratings. However, they are more difficult to fabricate on the sub-\SI{10}{\micro\meter} period scale for x-rays than binary TAIs discussed in this work. Furthermore, there is no benefit (in terms of visibility) in focusing on areas much smaller than the detector PSF. We conclude that the discussed 2D TAI design with $DC=1/3$ and $\varphi= 2/3\pi$ is an efficient and easy to fabricate x-ray phase array, serving the purpose of high-resolution phase contrast and dark-field imaging.   

\section{Experimental}

\begin{figure*}[h!]
\centering
\includegraphics[width=\linewidth]{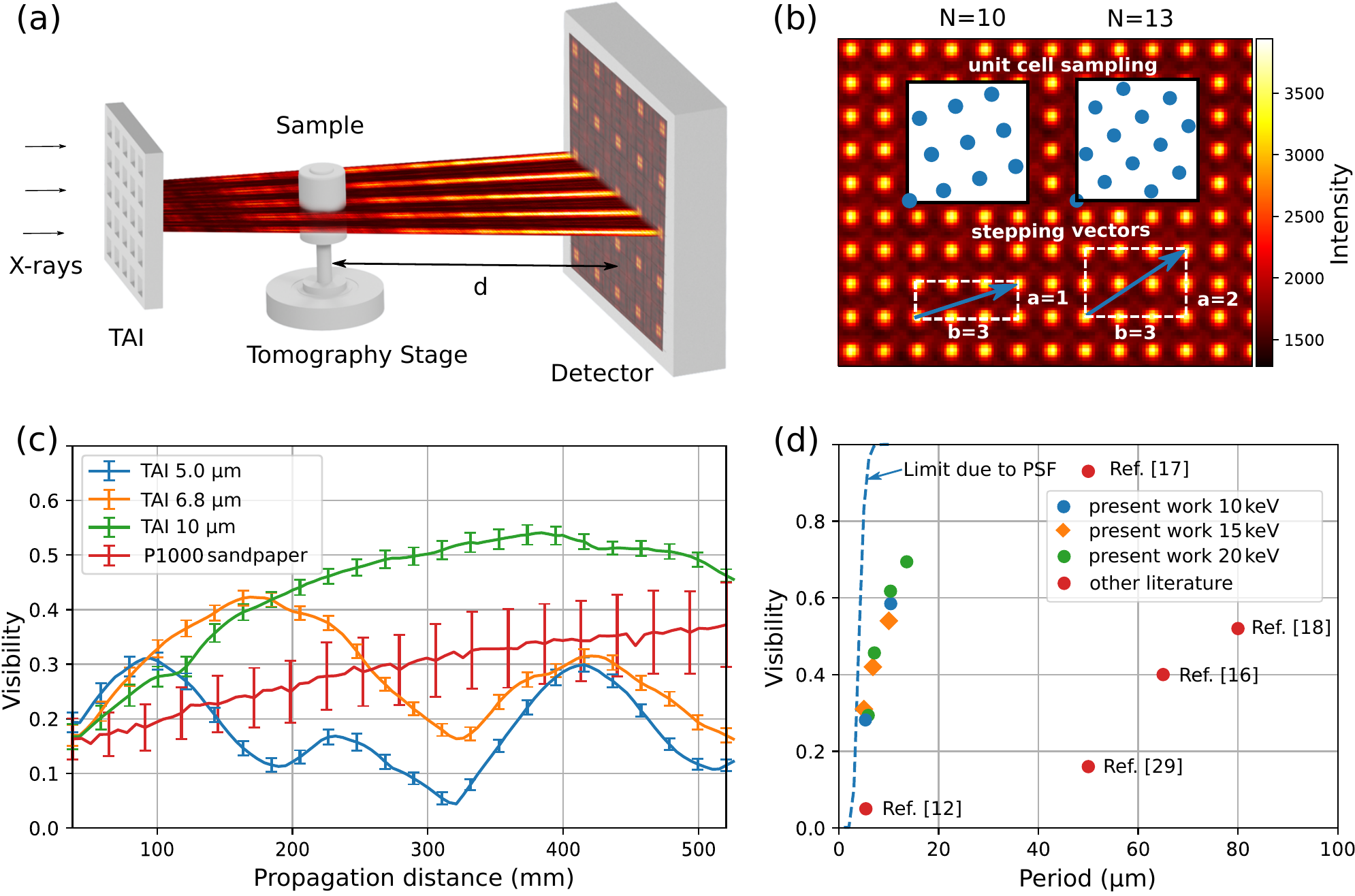}
\caption{(a) General setup of the imaging system illustrating the formation of the intensity modulation. (b) Measured intensity pattern with a high compression ratio in the background and illustration of the proposed 1D stepping scheme. The stepping vectors (blue arrows) represent the range and direction of the phase stepping. The sampling of the unit cell is exemplarily shown for both stepping vectors with $N=10$ and $N=13$ steps. (c) Distance-dependent visibility and its standard deviation (error bars every four points) for the TAIs and the P1000 diffuser extracted from the Talbot carpet scans at \SI{15}{\kilo \eV}. (d) Comparison of the evaluated TAIs in terms of visibility and period with recent literature.}
\label{fig:Fig2}
\end{figure*}

Multiple TAIs with periods of \SI{5.0}{\micro\meter}, \SI{6.8}{\micro\meter}, \SI{10.0}{\micro\meter} and \SI{13.6}{\micro\meter} with different heights adapted for energies of \SI{10}{\kilo\eV}, \SI{15}{\kilo\eV} and \SI{20}{\kilo\eV} were fabricated and evaluated for their diffractive properties using coherent x-ray synchrotron radiation at the P05 imaging beamline\cite{Greving2014, Wilde2016} operated by Helmholtz-Zentrum hereon at PETRA III at Deutsches Elektronen-Synchrotron (DESY), Hamburg, Germany. More details about the fabrication, setup parameters, and data processing are provided in Supplement 1. The general setup is shown in Figure \ref{fig:Fig2}(a), where incoming x-rays are modulated by the TAI, interact with the sample, and are then recorded by the detector. First, Talbot carpets (indicated by the colored layers in Figure \ref{fig:Fig2}(a)) were measured to confirm a higher compression ratio compared to conventional symmetric phase gratings and to find propagation distances with the best visibility for each TAI. One measured intensity modulation is shown in the background of Figure \ref{fig:Fig2}(b) for the TAI of \SI{6.8}{\micro\meter} period at \SI{15}{\kilo \eV}. Some images from the Talbot carpet scans are provided in Supplement 1 and compared with theoretical simulations. Similar scans have also been performed at \SI{10}{\kilo \eV} and \SI{20}{\kilo \eV} with the respective TAIs and some key parameters are listed in Supplement 1.

For comparison, a speckle pattern generated by a sheet of P1000 sandpaper representing a random phase modulator was also measured analogously to the Talbot carpet scans. The achieved visibility and its standard deviation according to equation \ref{eq:visibility} is plotted with increasing propagation distance in Figure \ref{fig:Fig2}(c) for all scans at \SI{15}{\kilo \eV} beam energy. To compare with recent literature demonstrating periodic x-ray DOEs discussed earlier, we plotted the peak performance of the different TAIs in \ref{fig:Fig2}(d). We only included research performed with single 2D gratings and modulators operated at synchrotron facilities for an appropriate comparison. The theoretical limit imposed by the PSF was calculated by convolving a periodic 2D square array of intensity points (resembling an ideally focusing modulator) with a Gaussian of $\sigma = \SI{1.5}{\micro\meter}$ estimating the blur of the used detector.

\begin{figure}[h]
\centering
\includegraphics[width=\linewidth]{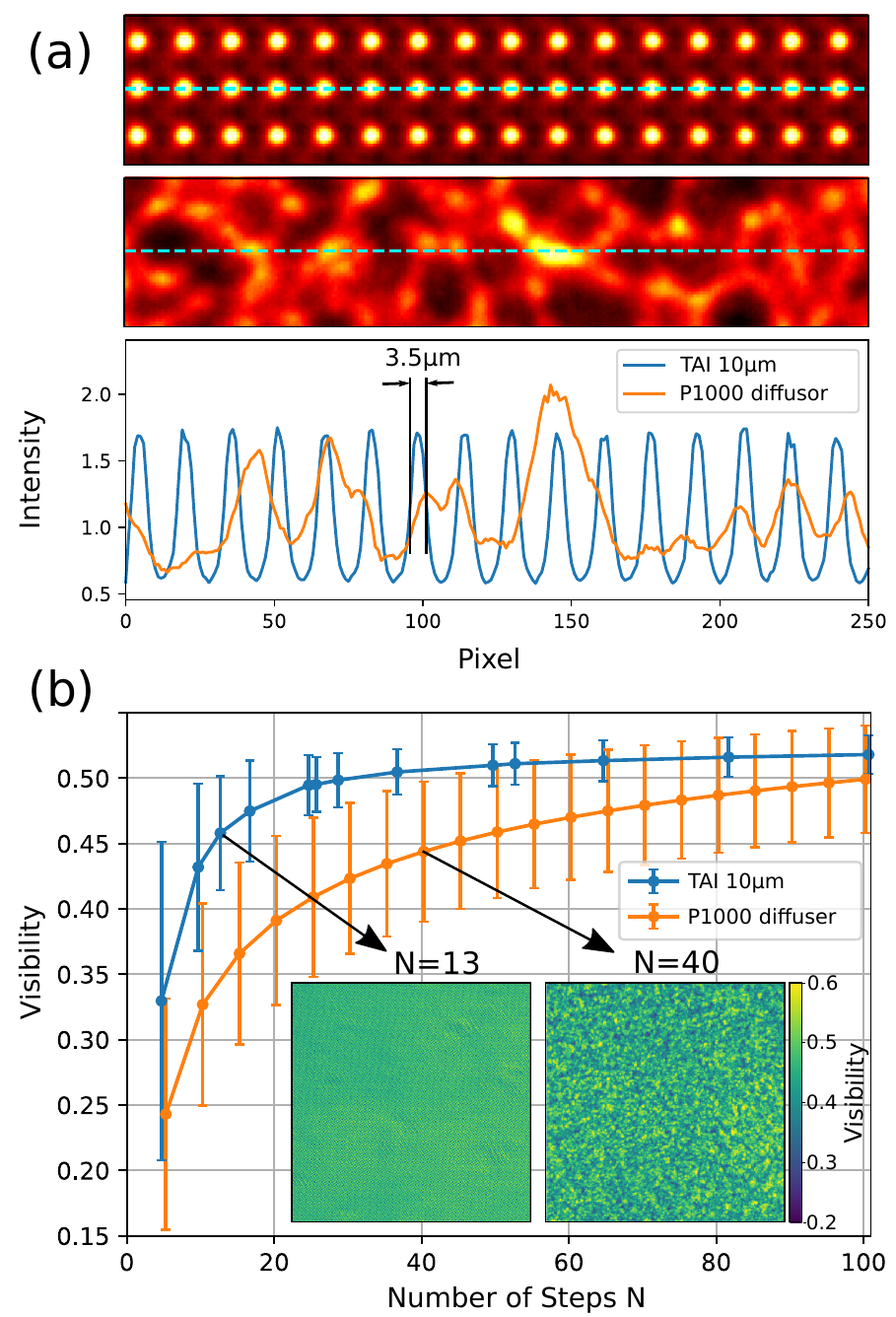}
\caption{(a) Comparison of intensity modulations created by the \SI{10}{\micro\meter} TAI and the P1000 diffuser with respective line plots illustrating the speckle sizes and densities. The images are normalized to the respective mean of the overall intensity values for a better comparison. (b) Comparison of visibilities and their standard deviations (error bars) with different numbers of steps for the TAI and the P1000 diffuser, including visibility maps for the TAI at $N=13$ and the P1000 at $N=40$. }
\label{fig:Fig3}
\end{figure}

For a better spatial resolution, a 1D stepping acquisition of the TAI for bi-directional phase sensitivity was evaluated and compared to a stepping procedure with the P1000 diffuser. The scheme, similarly applied before with absorption grids  \cite{Wen2010}, is illustrated in Figure \ref{fig:Fig2}(b). The stepping direction and range is chosen along a vector consisting of multiple unit cell vectors of the grating structure (e.g. $a=1$ and $b=3$) such that every unit cell is sampled uniformly by the periodic intensity maxima. This can be achieved by rotating the grating in an angle of $\arctan{(a/b)}$ to the stepping direction and performing $N = a^2+b^2 = 10$ steps. Such a homogeneous sampling can be achieved for different integers $a$ and $b$ when they are not coprime. It is noteworthy that homogeneous sampling does not necessarily have to be a quadratic lattice and can also be achieved with other rotation angles relative to the pixel matrix when the stepping range and step size can be precisely controlled. This approach, however, assumes that all foci sampling the unit cell have a very similar shape, which can be compromised by fabrication-related deficiencies. 
Figure \ref{fig:Fig3}(a) shows a comparison of the intensity pattern generated by the \SI{10}{\micro\meter} 2D TAI and the P1000 diffuser together with two line plots illustrating differences in the speckle densities and sizes. Stronger spots appear occasionally in the speckle pattern and would result in higher visibilities compared to the TAI when large pixel windows would be used for analysis according to \ref{eq:visibility}. For a realistic comparison close to the mode of imaging operation, stepping sets with different numbers of steps $N$ were composed from measured data and the overall visibility and its standard deviation were evaluated in every pixel. In the case of the TAI, the 1D stepping scheme discussed above was used and for the P1000 diffuser, a spiral stepping with an inter-step distance larger than the average speckle size was used to emulate a random stepping without repeated or very similar steps. The mean of the visibility and its standard deviation depending on the number of steps are shown in Figure \ref{fig:Fig3}(b). Exemplary visibility maps for the TAI with $N=13$ and P1000 with $N=40$ steps are plotted for comparison.

To demonstrate the imaging capabilities of the system, a sample consisting of silica particles with partially porous inner structures glued to a plastic micropipette tip was used. It was measured with the \SI{6.8}{\micro\meter} TAI placed \SI{170}{\milli\meter} from the detector at \SI{15}{\kilo\eV} beam energy. The sample was at $d= \SI{70}{\milli\meter}$ propagation distance to the detector and the TAI was stepped linearly according to the scheme discussed above. The acquired data was processed using Unified Modulated Pattern Analysis (UMPA) \cite{Zdora2017a} which is a flexible and robust algorithm suitable for both SBI as well as phase retrieval with periodic modulation patterns. More details about the processing and phase integration are provided in Supplement 1. Figure \ref{fig:Fig4}(a-d) shows different image channels acquired with $N = 17$ steps and processed with a window size of 5 pixels (\SI{0.64}{\micro\meter} effective pixel size). Some line plots in Figure \ref{fig:Fig4}(f) show selected small features of the silica particle from the dark-field and the phase channel together with inlets which were acquired with $N=25$ steps and processed with a window size of 3 pixels.     

An entire CT scan of a paraffin-embedded mouse artery (brachiocephalic) taken from an 'old' mouse (20 months) with atherosclerosis was performed at \SI{20}{\kilo\eV} acquiring $N=15$ steps per projection and an effective pixel size of \SI{0.91}{\micro\meter}. The CT scan was performed with a different detector configuration than the previous scans since a larger FoV was required. The sample was placed at $d = \SI{150}{\milli\meter}$ and 4001 projections were collected with intermediate flat fields. A specially adapted matching algorithm (described in detail in Supplement 1) was developed to find the most suitable flat field for every sample frame. The reconstruction was performed by filtered back-projection of the integrated phase images using a Ram-Lak filter. Figure \ref{fig:Fig5}(a) shows a 3D rendering of a section of the artery and an arrow is depicting a fissure in the vascular wall. The latter is shown in the respective slice (b) with a magnified view in (c). A line plot through the lamellar structure (see blue line in (b)) is plotted in (d) to estimate the achieved resolution. Further quantitative analysis related to the resolution is included in Supplement 1.

\section{Results and Discussion}

\subsection{Visibility-Distance Analysis}

As predicted by the simulation the evaluated TAIs create strongly modulated patterns near the fractional Talbot distances. In the case of the \SI{5}{\micro\meter} TAI, the first visibility peak appears in the vicinity of \SI{100}{\milli\meter} and the second around \SI{400}{\milli\meter} which corresponds to $1/6d_T$ and $2/3d_T$. Note that the highest visibility does not have to be necessary on the exact position of the fractional Talbot distance, since strong focusing occurs even before and after $1/6d_T$ as the simulations show. The $2/3d_T$ configuration can be used for high-sensitivity measurements as it delivers comparably high visibility at a quadrupled propagation distance than the modulation at $1/6d_T$ and has a significantly lower period than the \SI{6.8}{\micro\meter} TAI. The latter achieves higher visibility at $1/6d_T$ which is to be expected due to the PSF blur. The highest visibility of the TAIs for \SI{15}{\kilo\eV} is achieved by the \SI{10}{\micro\meter} TAI at \SI{380}{\milli\meter} which is close to the $1/6d_T$ modulation. The speckle visibility of P1000 sandpaper increases constantly with propagation distance, however, it is still always below the \SI{10}{\micro\meter} TAI and suffers by a much higher standard deviation. The TAIs show pivotal advantages for creating high contrast modulations at small periods and shorter propagation distances. As Figure \ref{fig:Fig2}(d) shows, the evaluated modulators are much closer to the theoretical limit in terms of the period-to-visibility ratio compared to other recent examples. Compared to \cite{Reich2018} the \SI{6.8}{\micro\meter} TAI achieves similar visibility at about $10\times$ smaller periods and enables therefore e.g. single-shot imaging at about one order of magnitude better resolution. Compared to e.g. \cite{Morgan2013} the configuration of the \SI{5}{\micro\meter} TAI gives a $6 \times$ higher visibility which allows to shorten the measurement time and lower the dose significantly. 

\begin{figure}[h!]
\centering
\includegraphics[width=\linewidth]{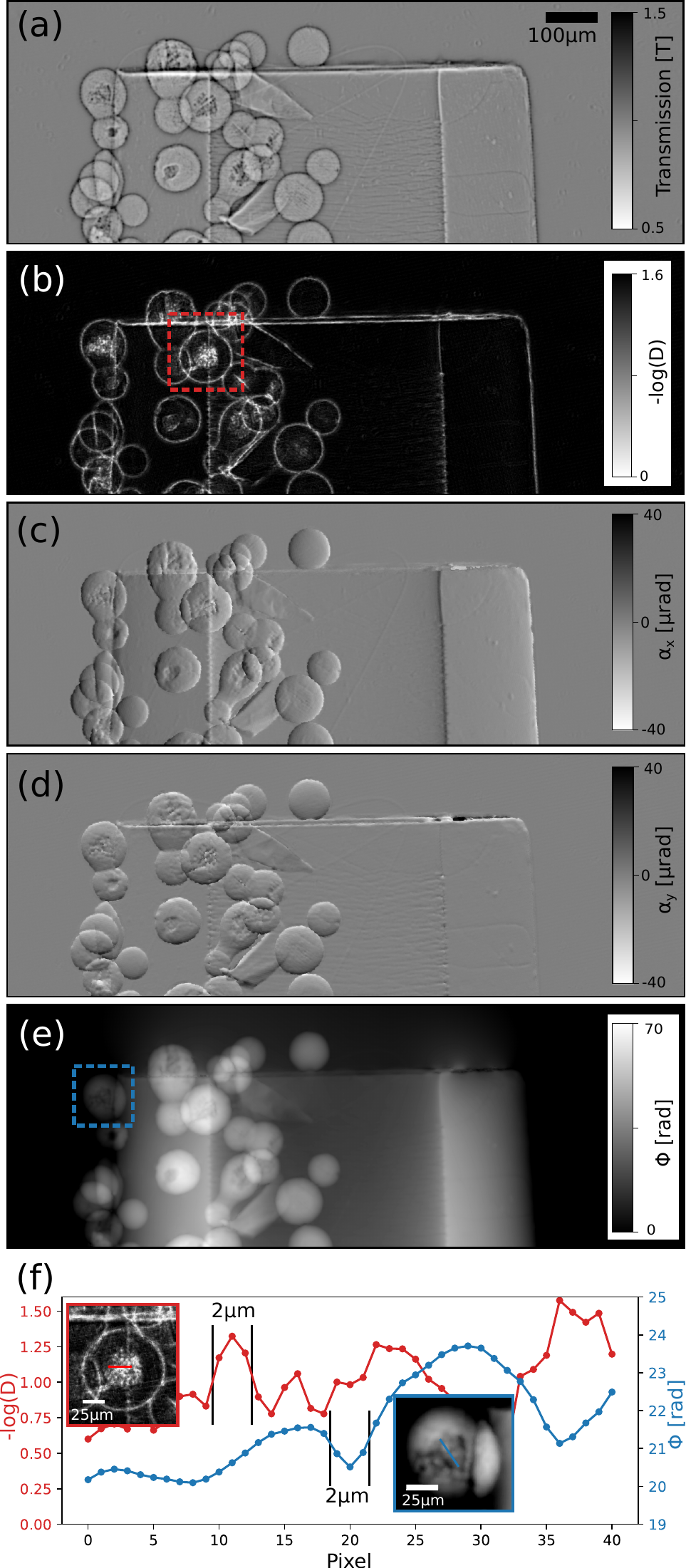}
\caption{Transmission image (a), dark-field image (b), differential phase contrast image in x-direction (c) and y-direction (d) and integrated phase image (e). Line plots of the dark-field and phase signal (f) from some selected features shown in the red and blue ROIs respectively.}  
\label{fig:Fig4}
\end{figure}

\begin{figure}[h!]
\centering
\includegraphics[width=\linewidth]{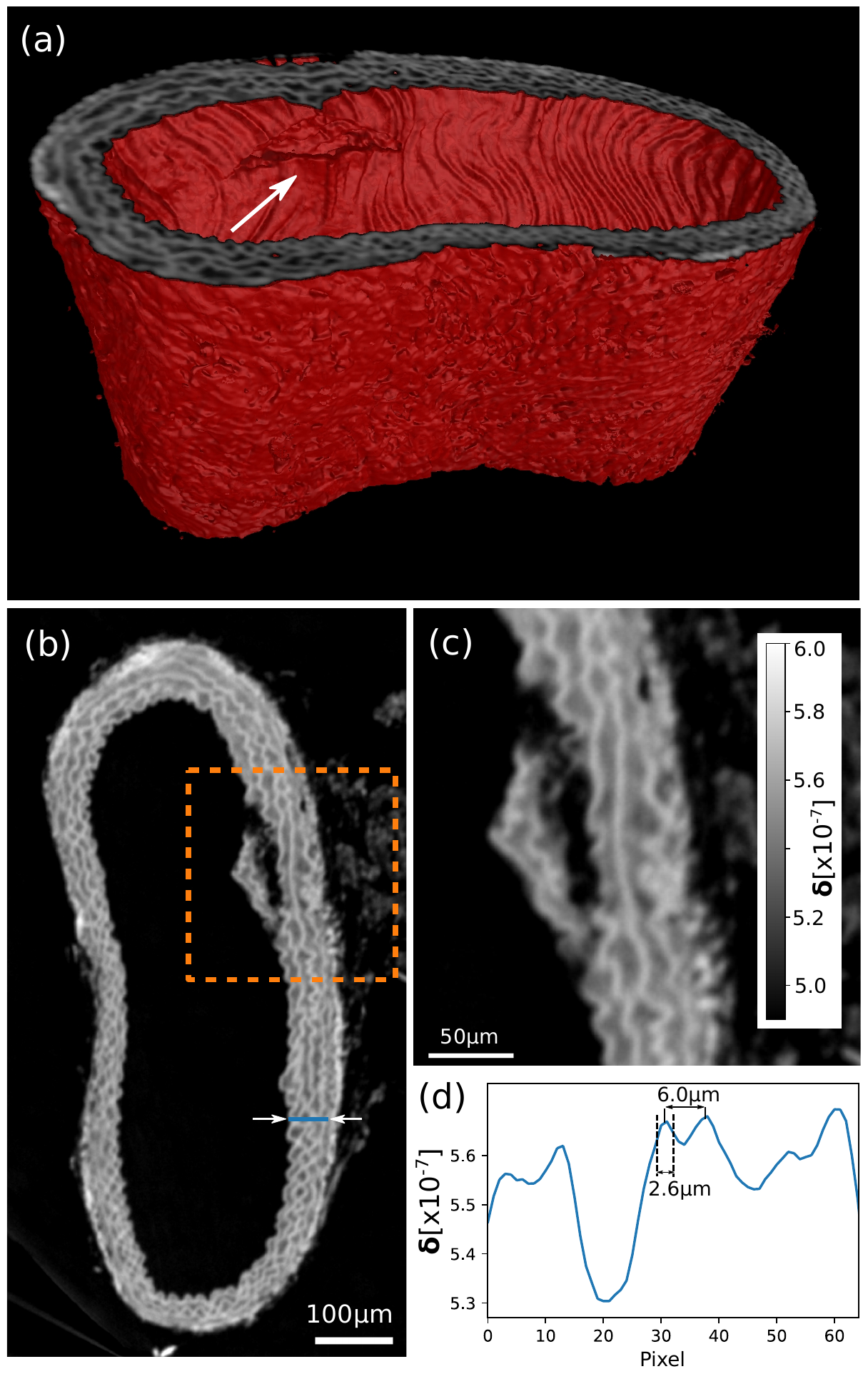}
\caption{(a) Rendering of a mouse aorta CT phase scan depicting a fissure in the vascular wall (white arrow) and a respective slice (b) showing its lamellar structure. (c) Magnified image of the ROI in (b). (d) Line plot (see blue line in (b)) showing that two lamellae distanced \SI{6}{\micro\meter} apart can be well resolved.}  
\label{fig:Fig5}
\end{figure}

\subsection{Comparison of TAI with P1000 Diffuser}

A comparison of periodic phase modulators with diffusers is not straightforward as they do not have a distinct period and the reached visibility either depends on the analysis window size $w$ or the number of steps when a pixel-by-pixel approach is used. Furthermore, it depends on the beam energy, coherence, and propagation distance. Still, a realistic evaluation of the achieved visibility in a phase stepping process with increasing $N$ shown in Figure \ref{fig:Fig3}(b) emphasizes the benefits of the TAI. The visibility, as well as its standard deviation, saturate after $N=20$ for the TAI, while the P1000 diffuser requires much more steps to achieve comparable results and its standard deviation decreases only slowly with increasing $N$. With $N=13$ an overall higher visibility with a lower standard deviation is achieved with the TAI than for the P1000 diffuser with $N=40$ steps. A detailed evaluation considering the intensity gradients of the modulation pattern, which are the key factor for a good phase sensitivity \cite{Zhou2016}, is given in Supplement 1. Using the TAI instead of the P1000 diffuser an improvement by a factor of 6 in dose efficiency is estimated. It is attributed to strong intensity gradients, their high density, as well as their periodic nature, which allows a highly efficient sampling.

\subsection{Projectional Imaging}

As shown in Figure \ref{fig:Fig4}, almost artifact-free images with a high resolution can be acquired with a relatively low number of steps. In the transmission image (a) features close to edges or grainy regions are distorted by edge enhancement effects (halos around spherical shapes) due to a long propagation distance to the detector. The dark-field image (b) shows characteristic enhancement of edges but also porous structures inside the spheres are well recognizable. Some faint periodic artifacts are present in the dark-field image probably caused by insufficient sampling ($N=17$ steps). They can be avoided by choosing a larger window size in UMPA processing, however, this will also reduce the resolution. Both differential-phase contrast images in x (c) and y (d) direction show artificially rough edges of the silicon spheres which are probably the result of incompatible phase sampling and window size. However, those artifacts are hardly recognizable in the integrated phase image (e). The sensitivity of the differential phase contrast images in both directions is similar ($\sigma_x = \SI{175}{\nano\radian}$, $\sigma_y = \SI{186}{\nano\radian}$), confirming an overall good sampling in both directions. (f) Line plots provided in Figure \ref{fig:Fig4}(f) show that features of about \SI{2}{\micro\meter} full width at half maximum (FWHM) can be resolved in both the dark-field as well as the phase image. That is close to the theoretical limit of UMPA phase retrieval imposed by the window size \cite{Zdora2018} (in this case $w = 3$) as well as the PSF of the detector. 

\subsection{Computed Tomography}

The phase CT scan of the unstained mouse artery demonstrates the potential for high-resolution, quantitative, non-destructive 3D virtual histology. The vessel was extracted from an aged animal with a pathological condition (atherosclerosis) and the slice in \ref{fig:Fig5}(b) shows in detail that there is a measurable fissure of the elastin fibers, commonly seen in frail, old animals and in humans. The contrast between the background (paraffin) and the bright elastin fibers is about \num{0.76E-7} and the background noise level in the paraffin matrix is \num{0.015E-7}. Hence, a contrast-to-noise ratio (CNR) of $>50$ between these soft matter components is achieved. A line plot of the reconstructed refractive index decrement $\delta$ values in (d) shows that two different lamellae distanced about \SI{6}{\micro\meter} from each other can still be well resolved and neighboring features with an FWHM of about \SI{2.6}{\micro\meter} are well distinguishable. Well defined edges quantifying the resolution at paraffin-tissue interfaces evaluated further in Supplement 1 also suggest a resolution of about \SI{3}{\micro\meter} achieved with $N=15$ phase steps per projection. A recent comparable work using state-of-the-art SBI achieved about \SI{8}{\micro\meter} resolution at $N=20$ phase steps at an energy of \SI{26.3}{\kilo\eV} \cite{Zdora2020}. Thus, we are approaching the limit imposed by the detector PSF and come close to the resolution of propagation-based phase imaging. The latter is unprecedented in resolution among non-ptychographic full-field techniques, however, is less sensitive to small density differences since it is based on the Laplacian of the phase. Furthermore, the most commonly used reconstruction algorithm \cite{Paganin2002} requires various assumptions about the sample, resulting in limited applications and difficulties for quantification. 

\section{Conclusion}

We designed and evaluated 2D TAIs with small periods for \numrange{10}{20}\si{\kilo\eV} x-ray beam energy. They create periodic foci with higher compression ratios and visibilities than conventional 2D phase gratings and have many advantages over absorption gratings or apertures, refractive micro-lens arrays, or random phase modulators used in SBI. The short periods also allow a finer and more efficient phase sampling for higher resolution and sensitivity. In this work we addressed the drawbacks of grating-based imaging (GBI) compared to SBI which are e.g. listed in \cite{Kashyap2016}. We avoided using absorptive elements and employed only one, thin phase modulator reducing the setup complexity. Furthermore, we reached bi-directional sensitivity with 1D linear phase stepping and achieved unprecedented resolution with a low number of phase steps. Similar to GBI, strong spatial modulations of $\delta$ (e.g. at edges, air bubbles) remain a problem, since they deteriorate the intensity pattern to a degree that it cannot be reasonably matched with the reference pattern. This could be addressed by tuning the sensitivity (sample-detector distance, beam energy, larger modulator period) or excluding the concerned pixels from the CT reconstruction using e.g. advanced iterative CT algorithms. Another possibility would be to apply ptychographic phase retrieval algorithms which have been shown beneficial for resolution with structured illumination \cite{Stockmar2013}.   
Future developments will include further optimization of the CT acquisition schemes to reduce the measurement time, radiation dose, and setup stability issues. In particular, a fly-scan CT with continuous sample rotation and frame rate at every phase step may improve the current protocol. Although a 1D stepping was successfully employed, a 2D stepping will simplify alignment and provide more flexibility for alternative sampling schemes. A detailed quantitative analysis of the absolute $\delta$ values as well as comparison to propagation-based phase tomography is ongoing and will be addressed in future work.  

The TAIs can be designed to operate efficiently in laboratory-based x-ray imaging systems with lower source coherence, shorter propagation distances, lower detector resolution, and higher x-ray energies. When SBI is performed with higher energies, strongly absorbing diffusers (e.g. steel wool \cite{Wang2016}) are used or multiple layers of sandpaper have to be stacked (e.g. up to 20 layers for \SI{65}{\kilo\eV} \cite{Berujon2017a}) to achieve decent speckle visibility. Periodic modulators like the TAIs proposed in this work can be designed for significantly higher energies on thin substrates. Current anisotropic silicon etching technology achieves aspect ratios beyond 1:20 making modulators with \SI{10}{\micro\meter} period possible for \SI{100}{\kilo\eV} and above on \SI{250}{\micro\meter} thick silicon substrates.

Beyond high-resolution or single-shot dynamic phase imaging, the discussed TAIs can be also used in wavefront sensing, x-ray optics characterization, adjustment and focusing of scintillator screens in 3 dimensions, or for recently demonstrated full-field structured illumination super-resolution x-ray microscopy \cite{Gunther2019, Mamyrbayev2020}. Using high power, laboratory-based x-ray sources with absorptive source gratings \cite{Pfeiffer2006e}, the discussed TAIs may be also used in medical Talbot-Lau-based imaging systems to gain bi-directional sensitivity and increased visibility with shorter setups compared to conventional binary symmetric phase gratings.

\section*{Funding Information}
Deutsche Forschungsgemeinschaft (DFG) Research Training Group GRK2274; European Research Council (ERC) grant agreement No 866026; British Heart Foundation (BHF) project PG/20/10010 

\section*{Acknowledgments}
Large parts of this research were carried out at PETRA III at DESY, a member of the Helmholtz Association (HGF). We acknowledge the support during the beam times by Jana Andrejewski, Fabio De Marco, Lev Ushakov, Elena Moralez, and Fabian Wilde. This research was supported in part through the Maxwell computational resources operated at DESY. We acknowledge the development and implementation of tailored detector systems for phase contrast imaging at the imaging beamlines P05 and P07 by members of the HZG tomography team. Pierre Thibault acknowledges funding from the European Research Council (ERC) under the European Union’s Horizon 2020 research and innovation program (grant agreement No 866026). 

\section*{Disclosures}
\noindent\textbf{Disclosures.} The authors declare no conflicts of interest.

\section*{Supplemental Documents}
For further methodical details see Supplement 1 attached at the end of the document.

{\small \bibliography{references}}  

\begin{thebibliography}{10}
\newcommand{\enquote}[1]{``#1''}

\bibitem{Fitzgerald2000}
R.~Fitzgerald, \enquote{{Phase‐Sensitive X‐Ray Imaging},}
  {\protect\JournalTitle{Physics Today}} \textbf{53}, 23--26 (2000).

\bibitem{Gureyev2009}
T.~E. Gureyev, S.~C. Mayo, D.~E. Myers, Y.~Nesterets, D.~M. Paganin, A.~Pogany,
  A.~W. Stevenson, and S.~W. Wilkins, \enquote{{Refracting R{\"{o}}ntgen's
  rays: Propagation-based x-ray phase contrast for biomedical imaging},}
  {\protect\JournalTitle{Journal of Applied Physics}} \textbf{105}, 102005
  (2009).

\bibitem{Davis1995}
T.~J. Davis, D.~Gao, T.~E. Gureyev, A.~W. Stevenson, and S.~W. Wilkins,
  \enquote{{Phase-contrast imaging of weakly absorbing materials using hard
  X-rays},} {\protect\JournalTitle{Nature}} \textbf{373}, 595--598 (1995).

\bibitem{Momose2003e}
A.~Momose, S.~Kawamoto, I.~Koyama, Y.~Hamaishi, K.~Takai, and Y.~Suzuki,
  \enquote{{Demonstration of X-Ray Talbot Interferometry},}
  {\protect\JournalTitle{Japanese Journal of Applied Physics}} \textbf{42},
  L866--L868 (2003).

\bibitem{Pfeiffer2006e}
F.~Pfeiffer, T.~Weitkamp, O.~Bunk, and C.~David, \enquote{{Phase retrieval and
  differential phase-contrast imaging with low-brilliance X-ray sources},}
  {\protect\JournalTitle{Nature Physics}} \textbf{2}, 258--261 (2006).

\bibitem{Olivo2007b}
A.~Olivo and R.~Speller, \enquote{{A coded-aperture technique allowing x-ray
  phase contrast imaging with conventional sources},}
  {\protect\JournalTitle{Applied Physics Letters}} \textbf{91}, 074106 (2007).

\bibitem{Morgan2011}
K.~S. Morgan, D.~M. Paganin, and K.~K.~W. Siu, \enquote{{Quantitative
  single-exposure x-ray phase contrast imaging using a single attenuation
  grid},} {\protect\JournalTitle{Optics Express}} \textbf{19}, 19781 (2011).

\bibitem{Berujon2012}
S.~B{\'{e}}rujon, E.~Ziegler, R.~Cerbino, and L.~Peverini,
  \enquote{{Two-Dimensional X-Ray Beam Phase Sensing},}
  {\protect\JournalTitle{Physical Review Letters}} \textbf{108}, 158102 (2012).

\bibitem{Morgan2012}
K.~S. Morgan, D.~M. Paganin, and K.~K.~W. Siu, \enquote{{X-ray phase imaging
  with a paper analyzer},} {\protect\JournalTitle{Applied Physics Letters}}
  \textbf{100}, 124102 (2012).

\bibitem{Zanette2014}
I.~Zanette, T.~Zhou, A.~Burvall, U.~Lundstr{\"{o}}m, D.~H. Larsson, M.~Zdora,
  P.~Thibault, F.~Pfeiffer, and H.~M. Hertz, \enquote{{Speckle-Based X-Ray
  Phase-Contrast and Dark-Field Imaging with a Laboratory Source},}
  {\protect\JournalTitle{Physical Review Letters}} \textbf{112}, 253903 (2014).

\bibitem{Wen2010}
H.~H. Wen, E.~E. Bennett, R.~Kopace, A.~F. Stein, and V.~Pai,
  \enquote{{Single-shot x-ray differential phase-contrast and diffraction
  imaging using two-dimensional transmission gratings},}
  {\protect\JournalTitle{Optics Letters}} \textbf{35}, 1932 (2010).

\bibitem{Morgan2013}
K.~S. Morgan, P.~Modregger, S.~C. Irvine, S.~Rutishauser, V.~A. Guzenko,
  M.~Stampanoni, and C.~David, \enquote{{A sensitive x-ray phase contrast
  technique for rapid imaging using a single phase grid analyzer},}
  {\protect\JournalTitle{Optics Letters}} \textbf{38}, 4605 (2013).

\bibitem{Berujon2017a}
S.~Berujon and E.~Ziegler, \enquote{{Near-field speckle-scanning-based x-ray
  tomography},} {\protect\JournalTitle{Physical Review A}} \textbf{95}, 063822
  (2017).

\bibitem{Zdora2017a}
M.-C. Zdora, P.~Thibault, T.~Zhou, F.~J. Koch, J.~Romell, S.~Sala, A.~Last,
  C.~Rau, and I.~Zanette, \enquote{{X-ray Phase-Contrast Imaging and Metrology
  through Unified Modulated Pattern Analysis},} {\protect\JournalTitle{Physical
  Review Letters}} \textbf{118}, 203903 (2017).

\bibitem{Zdora2018}
M.-C. Zdora, \enquote{{State of the Art of X-ray Speckle-Based Phase-Contrast
  and Dark-Field Imaging},} {\protect\JournalTitle{Journal of Imaging}}
  \textbf{4}, 60 (2018).

\bibitem{Reich2018}
S.~Reich, T.~{dos Santos Rolo}, A.~Letzel, T.~Baumbach, and A.~Plech,
  \enquote{{Scalable, large area compound array refractive lens for hard
  X-rays},} {\protect\JournalTitle{Applied Physics Letters}} \textbf{112},
  151903 (2018).

\bibitem{Rolo2018}
T.~{dos Santos Rolo}, S.~Reich, D.~Karpov, S.~Gasilov, D.~Kunka, E.~Fohtung,
  T.~Baumbach, and A.~Plech, \enquote{{A Shack-Hartmann Sensor for Single-Shot
  Multi-Contrast Imaging with Hard X-rays},} {\protect\JournalTitle{Applied
  Sciences}} \textbf{8}, 737 (2018).

\bibitem{Kagias2019}
M.~Kagias, Z.~Wang, M.~E. Birkbak, E.~Lauridsen, M.~Abis, G.~Lovric,
  K.~Jefimovs, and M.~Stampanoni, \enquote{{Diffractive small angle X-ray
  scattering imaging for anisotropic structures},}
  {\protect\JournalTitle{Nature Communications}} \textbf{10}, 5130 (2019).

\bibitem{Mamyrbayev2020}
T.~Mamyrbayev, A.~Opolka, A.~Ershov, J.~Gutekunst, P.~Meyer, K.~Ikematsu,
  A.~Momose, and A.~Last, \enquote{{Development of an Array of Compound
  Refractive Lenses for Sub-Pixel Resolution, Large Field of View, and
  Time-Saving in Scanning Hard X-ray Microscopy},}
  {\protect\JournalTitle{Applied Sciences}} \textbf{10}, 4132 (2020).

\bibitem{Zakharova2018}
M.~Zakharova, V.~Vlnieska, H.~Fornasier, M.~B{\"{o}}rner, T.~d.~S. Rolo,
  J.~Mohr, and D.~Kunka, \enquote{{Development and Characterization of
  Two-Dimensional Gratings for Single-Shot X-ray Phase-Contrast Imaging},}
  {\protect\JournalTitle{Applied Sciences}} \textbf{8}, 468 (2018).

\bibitem{Sato2011}
G.~Sato, T.~Kondoh, H.~Itoh, S.~Handa, K.~Yamaguchi, T.~Nakamura, K.~Nagai,
  C.~Ouchi, T.~Teshima, Y.~Setomoto, and T.~Den, \enquote{{Two-dimensional
  gratings-based phase-contrast imaging using a conventional x-ray tube},}
  {\protect\JournalTitle{Optics Letters}} \textbf{36}, 3551 (2011).

\bibitem{Itoh2011}
H.~Itoh, K.~Nagai, G.~Sato, K.~Yamaguchi, T.~Nakamura, T.~Kondoh, C.~Ouchi,
  T.~Teshima, Y.~Setomoto, and T.~Den, \enquote{{Two-dimensional grating-based
  X-ray phase-contrast imaging using Fourier transform phase retrieval},}
  {\protect\JournalTitle{Optics Express}} \textbf{19}, 3339 (2011).

\bibitem{Rutishauser2013b}
S.~Rutishauser, M.~Bednarzik, I.~Zanette, T.~Weitkamp, M.~B{\"{o}}rner,
  J.~Mohr, and C.~David, \enquote{{Fabrication of two-dimensional hard X-ray
  diffraction gratings},} {\protect\JournalTitle{Microelectronic Engineering}}
  \textbf{101}, 12--16 (2013).

\bibitem{Zanette2010}
I.~Zanette, T.~Weitkamp, T.~Donath, S.~Rutishauser, and C.~David,
  \enquote{{Two-Dimensional X-Ray Grating Interferometer},}
  {\protect\JournalTitle{Physical Review Letters}} \textbf{105}, 248102 (2010).

\bibitem{Suleski1997}
T.~J. Suleski, \enquote{{Generation of Lohmann images from binary-phase Talbot
  array illuminators},} {\protect\JournalTitle{Applied Optics}} \textbf{36},
  4686 (1997).

\bibitem{Arrizon1994}
V.~Arriz{\'{o}}n and J.~G. Ibarra, \enquote{{Trading visibility and opening
  ratio in Talbot arrays},} {\protect\JournalTitle{Optics Communications}}
  \textbf{112}, 271--277 (1994).

\bibitem{Szwaykowski1993}
P.~Szwaykowski and V.~Arrizon, \enquote{{Talbot array illuminator with
  multilevel phase gratings},} {\protect\JournalTitle{Applied Optics}}
  \textbf{32}, 1109 (1993).

\bibitem{Morimoto2015}
N.~Morimoto, S.~Fujino, Y.~Ito, A.~Yamazaki, I.~Sano, T.~Hosoi, H.~Watanabe,
  and T.~Shimura, \enquote{{Design and demonstration of phase gratings for 2D
  single grating interferometer},} {\protect\JournalTitle{Optics Express}}
  \textbf{23}, 29399 (2015).

\bibitem{Zakharova2019a}
M.~Zakharova, S.~Reich, A.~Mikhaylov, V.~Vlnieska, T.~d.~S. Rolo, A.~Plech, and
  D.~Kunka, \enquote{{Inverted Hartmann mask for single-shot phase-contrast
  x-ray imaging of dynamic processes},} {\protect\JournalTitle{Optics Letters}}
  \textbf{44}, 2306 (2019).

\bibitem{Rix2019}
K.~R. Rix, T.~Dreier, T.~Shen, and M.~Bech, \enquote{{Super-resolution x-ray
  phase-contrast and dark-field imaging with a single 2D grating and
  electromagnetic source stepping},} {\protect\JournalTitle{Physics in Medicine
  {\&} Biology}} \textbf{64}, 165009 (2019).

\bibitem{Mikhaylov2020}
A.~Mikhaylov, S.~Reich, M.~Zakharova, V.~Vlnieska, R.~Laptev, A.~Plech, and
  D.~Kunka, \enquote{{Shack–Hartmann wavefront sensors based on 2D refractive
  lens arrays and super-resolution multi-contrast X-ray imaging},}
  {\protect\JournalTitle{Journal of Synchrotron Radiation}} \textbf{27},
  788--795 (2020).

\bibitem{Yaroshenko2014c}
A.~Yaroshenko, M.~Bech, G.~Potdevin, A.~Malecki, T.~Biernath, J.~Wolf,
  A.~Tapfer, M.~Sch{\"{u}}ttler, J.~Meiser, D.~Kunka, M.~Amberger, J.~Mohr, and
  F.~Pfeiffer, \enquote{{Non-binary phase gratings for x-ray imaging with a
  compact Talbot interferometer},} {\protect\JournalTitle{Optics Express}}
  \textbf{22}, 547 (2014).

\bibitem{Greving2014}
I.~Greving, F.~Wilde, M.~Ogurreck, J.~Herzen, J.~U. Hammel, A.~Hipp,
  F.~Friedrich, L.~Lottermoser, T.~Dose, H.~Burmester, M.~Müller, and
  F.~Beckmann, \enquote{{P05 imaging beamline at PETRA III: first results},} in
  \emph{Developments in X-Ray Tomography IX,}  vol. 9212 S.~R. Stock, ed.,
  International Society for Optics and Photonics (SPIE, 2014), pp. 166 -- 173.

\bibitem{Wilde2016}
F.~Wilde, M.~Ogurreck, I.~Greving, J.~U. Hammel, F.~Beckmann, A.~Hipp,
  L.~Lottermoser, I.~Khokhriakov, P.~Lytaev, T.~Dose, H.~Burmester, M.~Müller,
  and A.~Schreyer, \enquote{Micro-ct at the imaging beamline p05 at petra iii,}
  {\protect\JournalTitle{AIP Conference Proceedings}} \textbf{1741}, 030035
  (2016).

\bibitem{Zhou2016}
T.~Zhou, M.-C. Zdora, I.~Zanette, J.~Romell, H.~M. Hertz, and A.~Burvall,
  \enquote{Noise analysis of speckle-based x-ray phase-contrast imaging,}
  {\protect\JournalTitle{Opt. Lett.}} \textbf{41}, 5490--5493 (2016).

\bibitem{Zdora2020}
M.-C. Zdora, P.~Thibault, W.~Kuo, V.~Fernandez, H.~Deyhle, J.~Vila-Comamala,
  M.~P. Olbinado, A.~Rack, P.~M. Lackie, O.~L. Katsamenis, M.~J. Lawson,
  V.~Kurtcuoglu, C.~Rau, F.~Pfeiffer, and I.~Zanette, \enquote{{X-ray phase
  tomography with near-field speckles for three-dimensional virtual
  histology},} {\protect\JournalTitle{Optica}} \textbf{7}, 1221 (2020).

\bibitem{Paganin2002}
D.~Paganin, S.~C. Mayo, T.~E. Gureyev, P.~R. Miller, and S.~W. Wilkins,
  \enquote{{Simultaneous phase and amplitude extraction from a single defocused
  image of a homogeneous object},} {\protect\JournalTitle{Journal of
  Microscopy}} \textbf{206}, 33--40 (2002).

\bibitem{Kashyap2016}
Y.~Kashyap, H.~Wang, and K.~Sawhney, \enquote{{Experimental comparison between
  speckle and grating-based imaging technique using synchrotron radiation
  X-rays},} {\protect\JournalTitle{Optics Express}} \textbf{24}, 18664 (2016).

\bibitem{Stockmar2013}
M.~Stockmar, P.~Cloetens, I.~Zanette, B.~Enders, M.~Dierolf, F.~Pfeiffer, and
  P.~Thibault, \enquote{{Near-field ptychography: phase retrieval for inline
  holography using a structured illumination},}
  {\protect\JournalTitle{Scientific Reports}} \textbf{3}, 1927 (2013).

\bibitem{Wang2016}
H.~Wang, Y.~Kashyap, B.~Cai, and K.~Sawhney, \enquote{{High energy X-ray phase
  and dark-field imaging using a random absorption mask},}
  {\protect\JournalTitle{Scientific Reports}} \textbf{6}, 30581 (2016).

\bibitem{Gunther2019}
B.~G{\"{u}}nther, L.~Hehn, C.~Jud, A.~Hipp, M.~Dierolf, and F.~Pfeiffer,
  \enquote{{Full-field structured-illumination super-resolution X-ray
  transmission microscopy},} {\protect\JournalTitle{Nature Communications}}
  \textbf{10}, 2494 (2019).

\end{thebibliography}


\begin{thebibliography}{1}
\newcommand{\enquote}[1]{``#1''}

\bibitem{Winthrop1965}
J.~T. Winthrop and C.~R. Worthington, \enquote{{Theory of Fresnel Images I
  Plane Periodic Objects in Monochromatic Light},}
  {\protect\JournalTitle{Journal of the Optical Society of America}}
  \textbf{55}, 373 (1965).

\bibitem{Lautner2017}
S.~Lautner, C.~Lenz, J.~U. Hammel, J.~Moosmann, M.~K{\"{u}}hn, M.~Caselle,
  M.~Vogelgesang, A.~Kopmann, and F.~Beckmann, \enquote{{Using SRuCT to define
  water transport capacity in Picea abies},}  (SPIE-Intl Soc Optical Eng,
  2017), p.~53.

\bibitem{Zdora2018}
M.-C. Zdora, \enquote{{State of the Art of X-ray Speckle-Based Phase-Contrast
  and Dark-Field Imaging},} {\protect\JournalTitle{Journal of Imaging}}
  \textbf{4}, 60 (2018).

\bibitem{Zdora2017a}
M.~C. Zdora, P.~Thibault, T.~Zhou, F.~J. Koch, J.~Romell, S.~Sala, A.~Last,
  C.~Rau, and I.~Zanette, \enquote{{X-ray Phase-Contrast Imaging and Metrology
  through Unified Modulated Pattern Analysis},} {\protect\JournalTitle{Physical
  Review Letters}} \textbf{118}, 203903 (2017).

\bibitem{Kottler2007}
C.~Kottler, C.~David, F.~Pfeiffer, and O.~Bunk, \enquote{{A two-directional
  approach for grating based differential phase contrast imaging using hard
  x-rays},} {\protect\JournalTitle{Optics Express}} \textbf{15}, 1175 (2007).

\bibitem{Zhou2016}
T.~Zhou, M.-C. Zdora, I.~Zanette, J.~Romell, H.~M. Hertz, and A.~Burvall,
  \enquote{{Noise analysis of speckle-based x-ray phase-contrast imaging},}
  {\protect\JournalTitle{Optics Letters}} \textbf{41}, 5490 (2016).

\bibitem{Modregger2007}
P.~Modregger, D.~L{\"{u}}bbert, P.~Sch{\"{a}}fer, and R.~K{\"{o}}hler,
  \enquote{{Spatial resolution in Bragg-magnified X-ray images as determined by
  Fourier analysis},} {\protect\JournalTitle{Physica Status Solidi (A)
  Applications and Materials Science}} \textbf{204}, 2746--2752 (2007).

\end{thebibliography}
\end{document}


\maketitle

\section{Detailed Materials and Methods}
\subsection{Simulation, Design, and Fabrication of TAIs}

\begin{figure}[htbp]
\centering
\includegraphics[width=\linewidth]{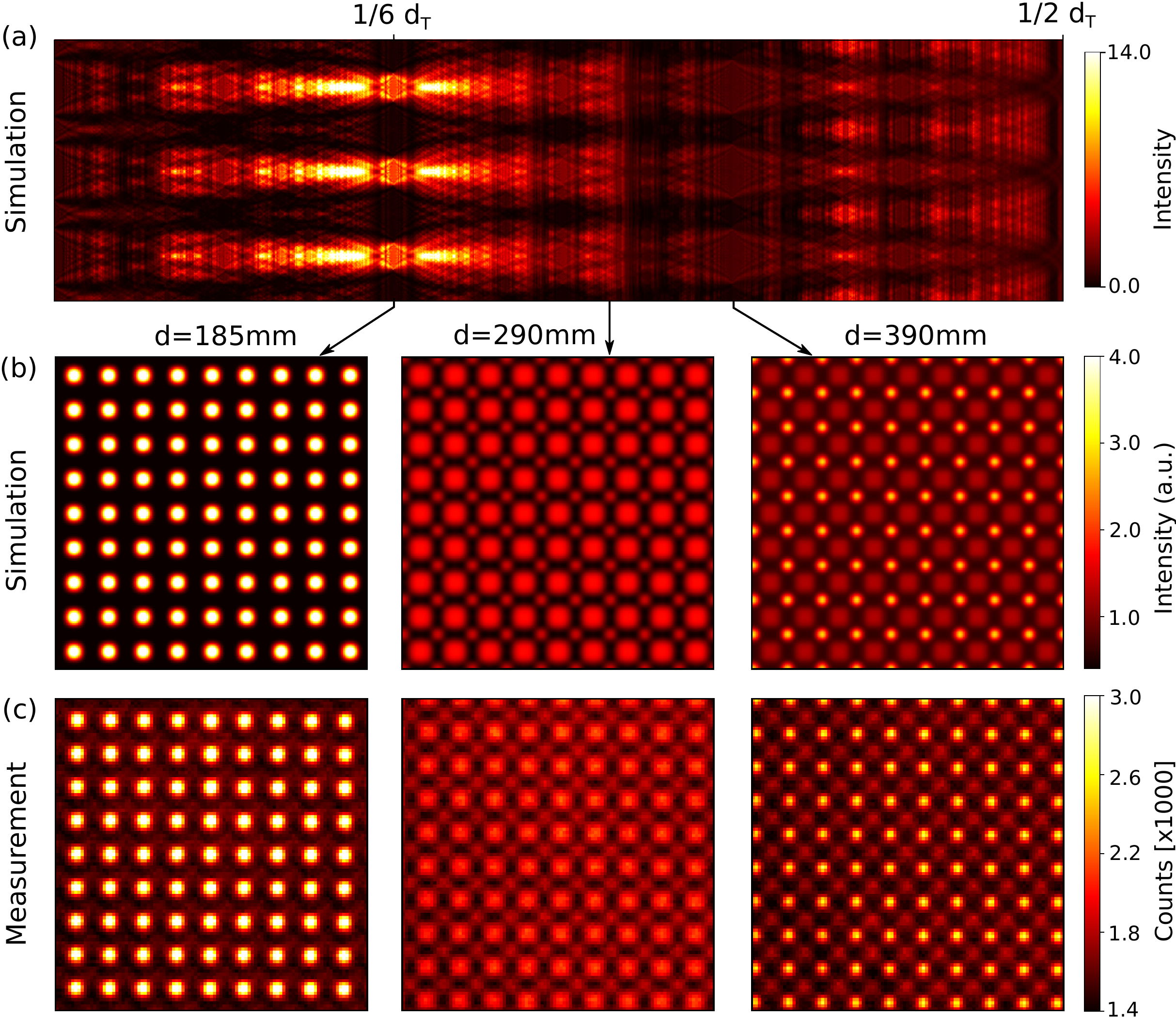}
\caption{(a) Simulation of the Talbot carpet for the TAI 6-15 at \SI{15}{\kilo\eV} (see Table 1) and (b) some Fresnel images at selected propagation distances after a convolution considering the detector PSF. (c) Acquired Fresnel images at the same propagation distances as in (b) agree well with the simulated data. }  
\label{fig:FigS1}
\end{figure}

\begin{figure}[h!]
\centering
\includegraphics[width=\linewidth]{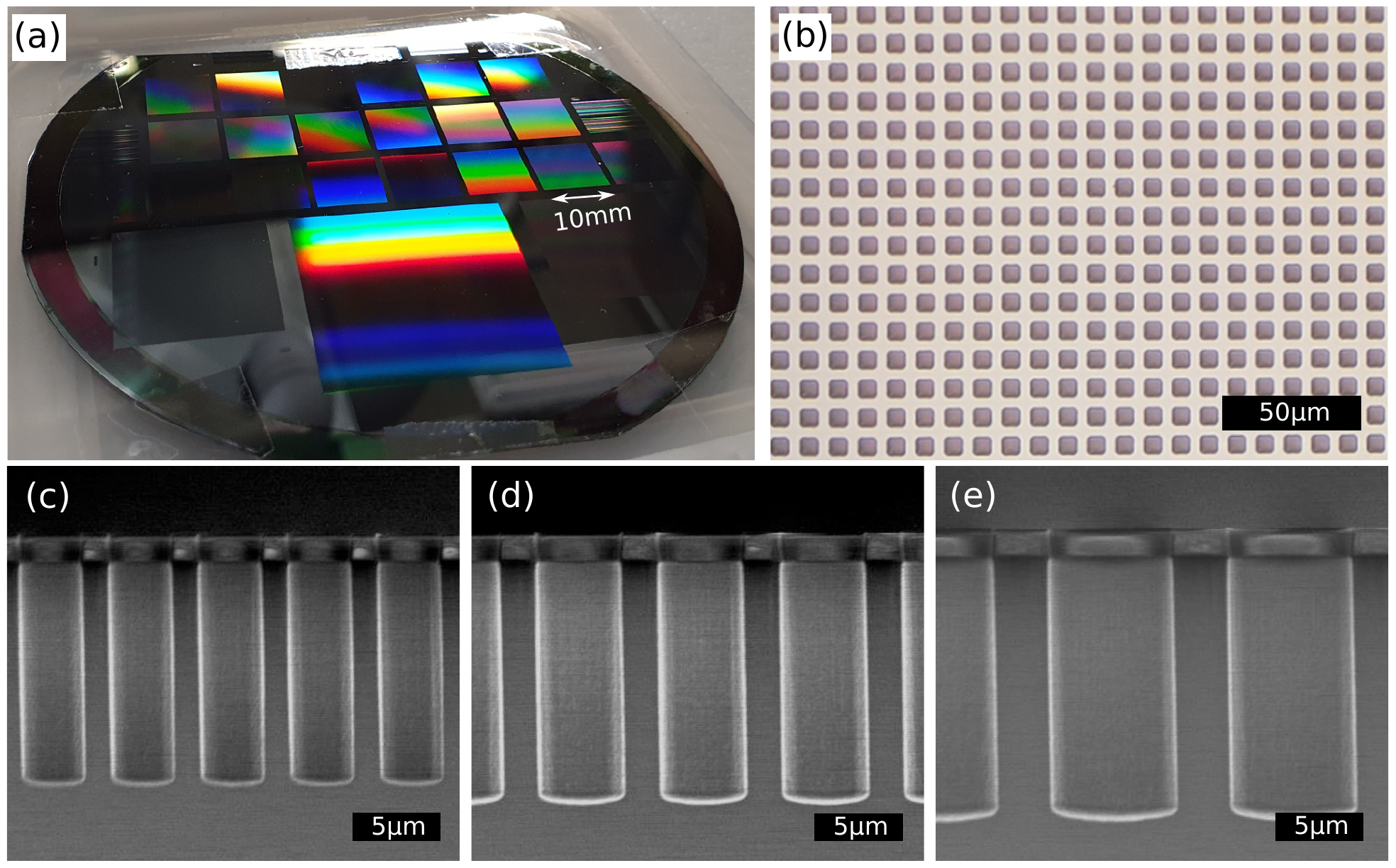}
\caption{(a) Multiple Talbot Array Illuminators (TAIs) with different periods on a \SI{100}{\milli\meter} silicon wafer. (b) Light microscopy image of the \SI{10}{\micro\meter} array. (c-e) SEM profile micrographs of the \SI{5}{\micro\meter} array (c), \SI{ 6.8}{\micro\meter} array (d) and \SI{10}{\micro\meter} array (e) etched simultaneously on the same wafer. Images (c-e) were kindly provided by 5microns GmbH (Illmenau, Germany).}  
\label{fig:FigS2}
\end{figure}

The simulation of coherent wave propagation was performed using the Fresnel diffraction formula,

\begin{equation}
\Phi(x,y,z) = \exp(ikz) F^{-1} \Biggl[F(\Phi_0) \exp \Biggl(-i\frac{k_x^2+k_y^2}{2k}z \Biggr)\Biggr],
\label{eq:refname1}
\end{equation}

where $\vec{k}$ is the wave vector, $ k = |\vec{k}|$ its modulus, $F$ denotes the Fourier transform ($F^{-1}$ is the inverse) and $\Phi_0 = \Phi(x,y, z=0)$ is the coherent wavefront created by the grating at $z=0$. The intensity $I$ measured by a detector is calculated by $I = \bigl|\Phi(x,y,z)\bigr|^2$ and is denoted as Fresnel image following the terminology from \cite{Winthrop1965}. Cross-sections of $I$ along the propagation direction are denoted as Talbot carpets. The calculations where performed with an area of $20 \times 20$ periods at \SI{100}{\nano\meter} resolution. We assumed the TAIs as pure phase objects as the attenuation by the modulator height is negligible at the respective beam energies. The Talbot carpet of the central 3 periods is shown in Figure \ref{fig:FigS1} (a). To obtain realistic visibility values the point spread function (PSF) of the detector had to be taken into account. For that, we convolved the simulated Fresnel images with a Gaussian kernel of $\sigma = \SI{0.7}{\micro\meter}$ in Figure \ref{fig:FigS1} (b) for a better comparison with the measured data (c). The design of the employed 2D phase modulators for x-ray applications was driven by multiple factors. They have to be (1) adapted to a good trade-off between visibility and period with a given detector PSF, (2) flux efficient using an x-ray transparent material, (3) resistant to high doses of radiation, and (4) easy to fabricate using current microprocessing technologies. Silicon was found to be a good choice given its widespread use in the semiconductor industry. The TAIs were fabricated according to our design specifications by 5microns GmbH (Illmenau, Germany) via UV lithography and deep reactive ion etching on \SI{250}{\micro\meter} thin silicon wafers. Multiple arrays of \SI{10}{\milli\meter}$\times$\SI{10}{\milli\meter} size were patterned and etched to different depths, which depend on the desired phase shifts at the respective beam energies. Table \ref{tab:TAI_paramterst} gives an overview of the different TAIs including the design parameters and the measured values for the height and the duty cycle based on scanning electron microscopy (SEM) images of the profiles. Some images of the gratings and micrographs of the profiles are shown in Figure \ref{fig:FigS2}.

\begin{table}[htbp]
\centering
\caption{\bf Fabrication-related parameters of the evaluated TAIs.}
\begin{tabular}{ccccc}
\hline
Modulator & Period[\si{\micro\meter}] & Energy[\si{\kilo \eV}] & Height [\si{\micro\meter}] (design/actual) & Duty Cycle (design/actual) \\ 
\hline
TAI 5-10  &   5.0   & 10 & 8.5 / 8.1  & 0.33 / 0.31     \\
TAI 10-10 &  10.0  & 10 & 8.5 / 9.1  & 0.33 / 0.31     \\
\hline
TAI 5-15  &   5.0   & 15 & 12.8 / 12.5  &  0.33 / 0.30     \\
TAI 6-15  &   6.8   & 15 & 12.8 / 13.5  &  0.33 / 0.29     \\
TAI 10-15 &   10.0  & 15 & 12.8 / 14.9  &  0.33 / 0.31     \\
\hline
TAI 5-20   &  5.0  & 20 & 17.1 / 15.5 &   0.33 / 0.27      \\
TAI 6-20   &  6.8  & 20 & 17.1 / 17.1 &   0.33 / 0.29      \\
TAI 10-20  &  10.0 & 20 & 17.1 / 18.7 &   0.33 / 0.30      \\
TAI 13-20  &  13.6 & 20 & 17.1 / 19.9 &   0.33 / 0.31      \\

\end{tabular}
  \label{tab:TAI_paramterst}
\end{table}

\begin{table}[htbp]
\centering
\caption{\bf Experimental parameters for the evaluated phase modulators. *For the P1000 diffuser the average speckle size is given instead of the FWHM focus size.}
\begin{tabular}{ccccc}
\hline
Modulator & Energy[\si{\kilo\eV}] & Analysis window [px] & FWHM Foci [\si{\micro\meter}] & Maximum Visibility   \\ 
\hline
TAI 5-10   & 10 & 8  & 1.9  & 0.27     \\
TAI 10-10  & 10 & 16 & 4.2  & 0.58     \\
\hline
TAI 5-15   & 15 & 8  &  2.0  & 0.31    \\
TAI 6-15   & 15 & 12 &  2.5  & 0.42    \\
TAI 10-15  & 15 & 16 &  3.5  & 0.54    \\
\hline
P 1000     & 15 & 18 &  6.9* & 0.36   \\
\hline
TAI  5-20  & 20 &  8 &  2.4   & 0.29   \\
TAI  6-20  & 20 & 12 &  2.8   & 0.45   \\
TAI 10-20  & 20 & 16 &  3.9   & 0.61   \\
TAI 13-20  & 20 & 22 &  5.3   & 0.68   \\

\end{tabular}
  \label{tab:TAI_result}
\end{table}

\subsection{Measurement of Talbot Carpets}

Multiple Talbot carpet scans were performed at the micro-tomography end-station of the P05 imaging beamline (IBL) at PETRA III at DESY. The radiation generated by an undulator source about \SI{86}{\meter} before its exit point to the experimental chamber is monochromatized by a double crystal monochromator. The detector was a \SI{50}{\micro\meter} thick lutetium-aluminium garnet (LuAG) scintillator optically coupled ($10\times$ magnification) to a CMOS camera (CMOSIS CMV20000 chip) resulting in an effective pixel size of \SI{0.64}{\micro\meter}\cite{Lautner2017}. Images were taken every \SI{5}{\milli\meter} from \SI{40}{\milli\meter} to \SI{540}{\milli\meter} distance to the detection plane with an exposure time of \SI{200}{\milli\second}. All measured TAIs with the respective periods and energies are given in Table \ref{tab:TAI_paramterst}. Figure \ref{fig:FigS2}(c) shows exemplary ROIs of the measured Fresnel images in different propagation distances acquired with the TAI 6-15 at \SI{15}{\kilo\eV}. The patterns match the simulated data (\ref{fig:FigS2}(b)) well, except that their contrast is somewhat lower due to the detector PSF.

\subsection{Characterization of the P1000 Diffuser}
\begin{figure}[h!]
\centering
\includegraphics[width=\linewidth]{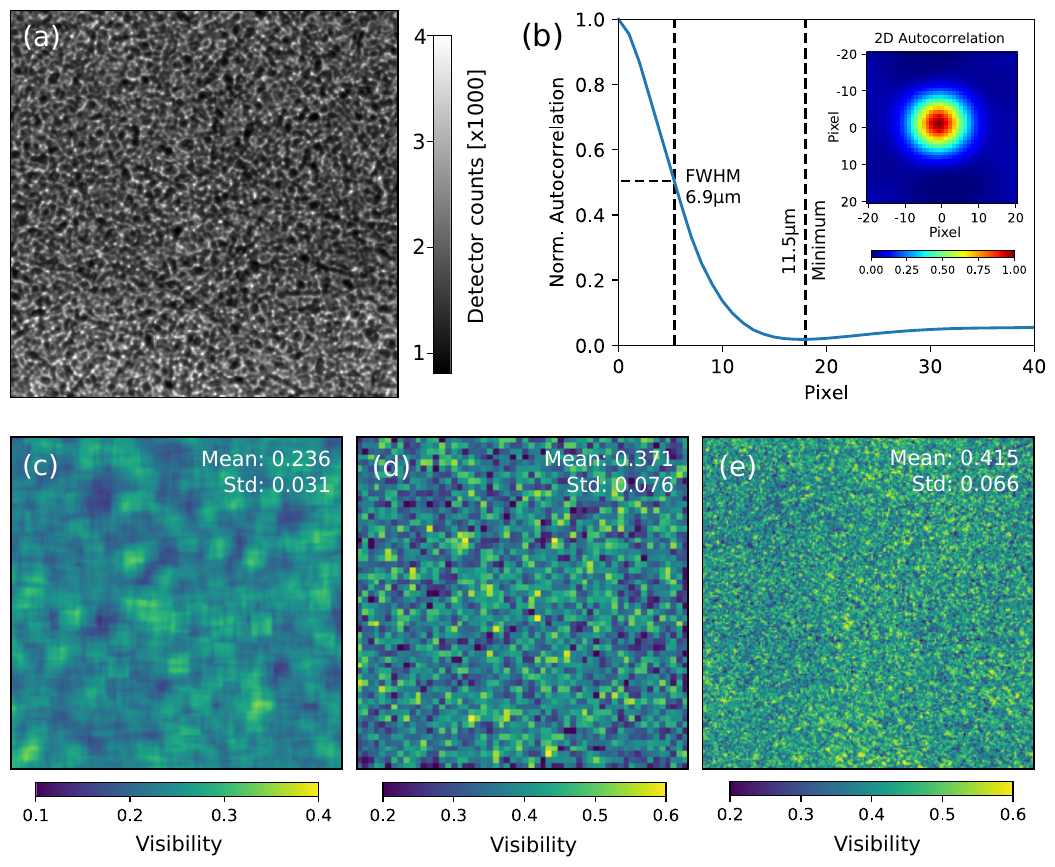}
\caption{(a) Measured speckle pattern ($1000 \times 1000$ pixel central ROI) at \SI{380}{\milli\meter} propagation distance. (b) Azimuthally-averaged plot of the 2D autocorrelation (seen in the inlet) of the speckle pattern suggesting an average speckle size of \SI{6.9}{\micro\meter}. (c) Visibility map calculated by convention from equation \ref{eq:visstd} with a window size of $50 \times 50$ pixels and (d) calculated by equation \ref{eq:visminmax}. (e) Visibility map calculated pixel-by-pixel from a spiral stepping with $N = 40$.}  
\label{fig:FigS3}
\end{figure}

A sheet of P1000 sandpaper (representative for a diffuser used in SBI) was measured with the same protocol to study and compare its speckle contrast with the evaluated TAIs. Figure \ref{fig:FigS3}(a) shows a $1000 \times 1000$ pixel ROI from the center of the speckle pattern at \SI{15}{\kilo\eV} and a propagation distance of \SI{380}{\milli\meter}. An azimuthally averaged and normalized plot of the 2D autocorrelation function of the speckle pattern is given in Figure \ref{fig:FigS3}(b). The half-width is at about 5.4 pixels resulting in a full width at half maximum (FWHM) of about \SI{6.9}{\micro\meter} which is a good measure of the average speckle size. The minimum at about 18 pixels (\SI{11.5}{\micro\meter}) is an average distance at which a maximal intensity modulation is expectable and was used as window size in the visibility determination. The 2D autocorrelation is also given as an inlet in Figure \ref{fig:FigS3}(b) showing that the beam coherence in both directions was high enough to not impair the speckle contrast at that propagation distance. Speckle patterns are characterized by different visibility conventions which are usually not comparable to each other \cite{Zdora2018}. Figure 3(c) shows the visibility map determined in a window of $50 \times 50$ pixels by standard deviation $\sigma_w$:

\begin{equation}
V_{std} = \frac{\sigma_{w}}{\overline{I_{w}}},
\label{eq:visstd}
\end{equation}

where $\overline{I_{w}}$ is the mean of the intensity values in the window. Figure \ref{fig:FigS3}(d) shows a visibility map determined by the minimal ($I_{\mathrm{min}}$) and maximal ($I_{\mathrm{max}}$) intensity pixel values in a window size of 18 pixels by

\begin{equation}
V = \frac{I_{\mathrm{max}}-I_{\mathrm{min}}}{I_{\mathrm{max}}+I_{\mathrm{min}}}.
\label{eq:visminmax}
\end{equation}

This convention is also used for the comparison with the TAIs at respective window sizes. Figure \ref{fig:FigS3}(e) shows a visibility map extracted pixel-by-pixel by formula \ref{eq:visminmax} from a stepping with $N=40$ random steps. The averages and standard deviations of the respective visibility arrays are given in the right corners of the maps respectively. Note that Figure \ref{fig:FigS3}(c) has a different leveling of the colormap. 

\subsection{Visibility Comparison of 10µm TAI and P1000 Diffuser}

All measured Fresnel images from the Talbot carpet scans were corrected with dark frames for read-out noise and strong pixel outliers (defective pixels) were replaced by its neighbors using a selective median filter. The visibility was extracted according to equation \ref{eq:visminmax} from the pixel intensity values in a window size (see Table \ref{tab:TAI_result}) similar to the period of each TAI. For the P1000 diffuser, the window size was chosen according to the first minimum of the azimuthally averaged autocorrelation function as described above. To evaluate the standard deviation of the visibility this analysis was performed in a $1000 \times 1000$ pixel array. The average visibility and its standard deviation represented by the error bars (only every 4th error bar was plotted for clarity) for the \SI{15}{\kilo\eV} measurements are plotted in Figure 2(c) of the main article.

To compare the realistic performance of the proposed 1D stepping scheme with a random stepping of the P1000 modulator, we composed stepping sets with different numbers of steps and calculated the resulting pixel-by-pixel visibility according to equation \ref{eq:visminmax}. Note that we chose equal propagation distances for the comparison of both modulators because different distances would result in incomparable flux at the detector and source coherence requirements when using a laboratory-based cone-beam system. If this aspect is irrelevant, a longer propagation distance could be used for the speckle pattern to reach a higher visibility. However, even at a measured propagation distance of \SI{1000}{\milli\meter} the speckle visibility of P1000 was 0.42, which is significantly lower than that of the TAI at \SI{380}{\milli\meter} (0.54). 

Since a precise 2D stepping stage for the phase modulators was not available during the experiments we used the following procedure to assemble sets of steppings with different $N$. For P1000 we shifted the measured speckle pattern on a spiral trajectory with a step size larger than the speckle size to emulate a random stepping. For the 2D TAI, the measured frames were rotated to an angle $\alpha = \arctan(a/b)$ and shifted digitally with sub-pixel interpolation to the respective positions, covering the range $a^2+b^2$ in $N$ steps according to the scheme described in the main article. For all stepping sets, the visibility was calculated pixel by pixel and its average and standard deviation were plotted versus the number of steps in Figure 3(c) of the main work. It should be noted that the sandpaper used for our comparison is not optimized for that energy and should not be considered as the ideal case of SBI. In general, a rigorous comparison between random and periodic phase modulators is difficult, since it requires finding the best random modulator for the set of parameters that a periodic modulator (e.g. TAI, grating, refractive lens array) is designed for. In our case, the P1000 sandpaper achieves relatively high visibility and is comparable to recent SBI literature. 

\subsection{Imaging of Silica Particles}

A sample consisting of porous silica particles glued on a plastic micropipette tip was used to test the resolution in both the phase and the dark-field modality. The \SI{6.8}{\micro\meter} TAI was operated at a distance of \SI{170}{\milli\meter} to the detector (same as for the Talbot carpet scans). The sample-detector distance was \SI{70}{\milli\meter}. The TAI was slightly tilted to the detector pixel matrix and for different numbers of steps, the stepping distance was chosen accordingly to achieve a homogeneous sampling. Additionally, multiple sets of flat fields were collected with the same scheme. 
Due to some possible experimental deficiencies like drifts of the beam, sample stage, or phase modulator mount as well as imprecise reproducibility of phase steps a digital matching of flat fields had to be performed prior to phase retrieval. For that, every sample frame was matched with the best acquired flat field using an image similarity algorithm on a sample-free area of the projection. To increase the precision the best flat field was further shifted digitally with sub-pixel interpolation in both directions to achieve optimal registration with the intensity pattern on the sample-free area. The phase retrieval was performed by Unified Modulated Pattern Analysis (UMPA) \cite{Zdora2017a} with different numbers of phase steps and window sizes. Note that since the angle of the stepping direction to the grating pattern could not be controlled precisely, there are only a few configurations for $N$ that produce a satisfactory sampling. Some of these configurations and the pixel-wise flat-field visibility and sensitivity are listed in Table \ref{tab:Sens}.
Both directional differential phase contrast images were used for a Fourier transform-based phase integration procedure described in detail in \cite{Kottler2007}.

\subsection{Specimen Preparation}

A 17-month-old C57BL6/J mouse purchased from Charles River UK was transduced with AAV8-PCSK9 (\num{6e12} vg/mouse i.v.) and one week later was started on a high-fat Western type diet, (829100, Special Diet Services, UK) for 12 weeks to induce atherosclerosis. At 20 months of age, the mouse was terminally anesthetized with pentobarbital and perfusion fixed via the left ventricle, first with saline (\SI{2}{\milli\litre}) to remove the blood and then with 10\% v/v buffered formalin (\SI{5}{\milli\litre}). Directly after this, the entire aortic tree (aortic arch, surrounding vessels, and heart) was dissected as a single entity and stored in fixation solution at \SI{4}{\celsius}. The brachiocephalic artery was embedded in paraffin wax for the CT scan. All experiments were approved by the local review board and the UK Home Office under license P5395C858. All mice were housed under standard light and dark conditions and had unlimited access to diet and water at all times. 

\subsection{X-ray phase CT of a Murine Artery}

The murine artery embedded in paraffin wax features fine lamellar structures in the tunica media, well suitable for a good resolution and soft-tissue contrast benchmark. Since the sample was larger than the previous sample, the detector configuration had to be adapted for a larger FoV. The camera (Ximea CB500MG, \SI{4.6}{\micro\meter} pixel size) was coupled to a $5\times$ optical magnification lens system. A \SI{100}{\micro\meter} Cadmium tungstate scintillator screen with a lower resolution but higher photon efficiency was focused at \SI{20}{\kilo\eV} yielding an effective pixel size of \SI{0.91}{\micro\meter}. The FoV in this configuration was $\SI{7}{\milli\meter}\times \SI{2.5}{\milli\meter}$.
The sample was mounted on the tomography stage at a propagation distance of $d = \SI{150}{\milli\meter}$ to the detector. At every angle, 15 phase steps were performed with an exposure time of \SI{200}{\milli\second} each and a total of 4001 projections were acquired over \SI{180}{\degree} sample rotation.   
To avoid ring artifacts in the later reconstruction the sample was laterally shifted in-between projections. Sets of flat fields were acquired multiple times to account for setup-related drifts and instabilities. The total scan time (about 5h) was comparably long due to experimental overhead (mechanical stepping, sample shifts, sub-optimal frame acquisition parameters) and can be significantly reduced, especially when a smaller FoV and consequently lower numbers of projections are required. Image processing was performed as previously described by a matching algorithm for flat fields and subsequent UMPA phase retrieval with a window size of 3 pixels. The CT reconstruction was performed via filtered back-projection (Ram-Lak filter) of the integrated phase images using the software X-Aid FDK Reconstruction Suite 2020.10.2 (Mitos GmbH, Garching, Germany). The reached sensitivity and visibility of the CT scan are given in Table \ref{tab:Sens}.

\section{Phase sensitivity, Period and Visibility}

The angular sensitivity of an imaging system refers to the ability of the setup to resolve small-angle deflections of the beam caused by refraction or scattering in the sample. The differential phase signal is directly related to the refraction angle $\alpha$ via \cite{Zdora2018}:

\begin{equation}
\begin{pmatrix} \partial\Phi / \partial x \\  \partial\Phi / \partial y \end{pmatrix} = \frac{2\pi}{\lambda} \begin{pmatrix} \alpha_x \\ \alpha_y \end{pmatrix}. 
\label{eq:dPhi}
\end{equation}

To measure $\alpha$ the deflection of the intensity modulation must be resolved. Using the small-angle approximation it translates to

\begin{equation}
\begin{pmatrix} \alpha_x \\ \alpha_y \end{pmatrix} = \frac{p_{\mathrm{eff}}}{d} \begin{pmatrix} u_x \\ u_y \end{pmatrix}, 
\label{eq:Alpha}
\end{equation}

where $d$ is the distance from sample (placed after phase modulator) to detector, and $(u_x,u_y)$ is the displacement of the pattern in units of the effective pixel size $p_{\mathrm{eff}}$.

\begin{figure}[htbp]
\centering
\includegraphics[width=\linewidth]{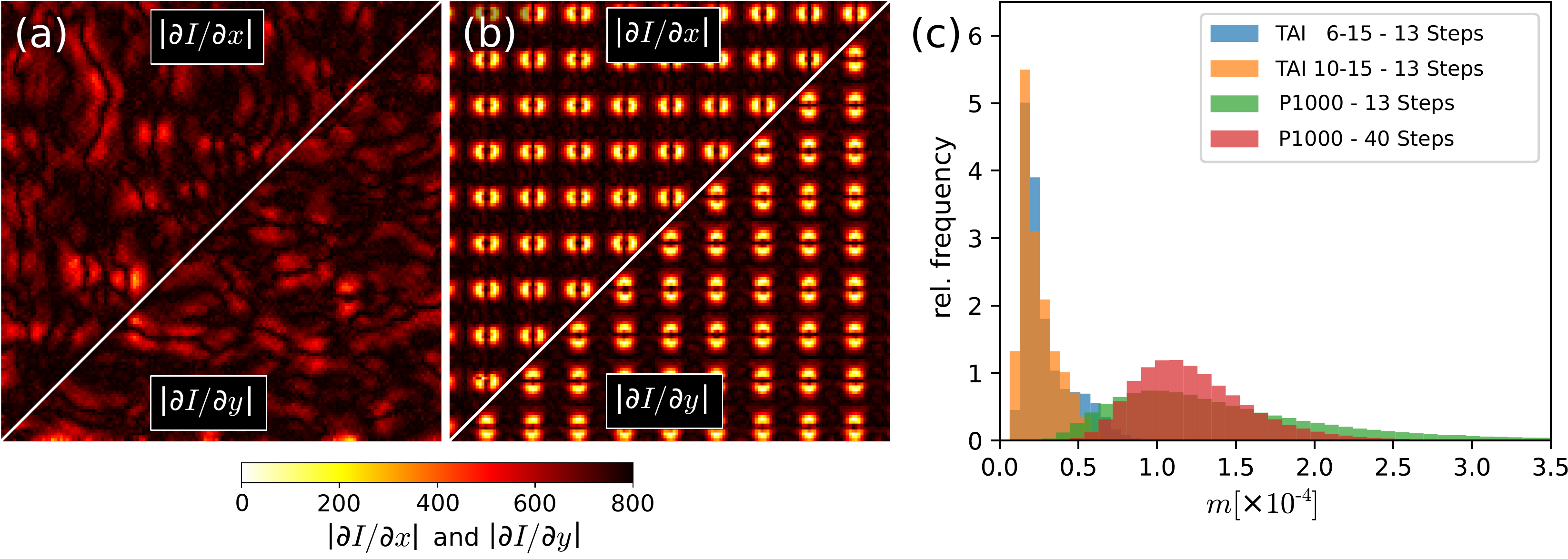}
\caption{Modulus gradients of the P1000 speckle pattern (a) and the TAI 10-15 modulator (b) in a 150 $\times$ 150 pixel ROI. (c) Histograms of $m$ calculated from a 500 $\times$ 500 pixel ROI for different modulators and phase stepping scenarios.}  
\label{fig:FigSGradient}
\end{figure}

In single-shot X-ray speckle-tracking, small window patches from the sample and the reference scan are compared to calculate its displacement reaching a resolution in the range of the average speckle size \cite{Zdora2018}. Various techniques allow increasing resolution and sensitivity by stepping the modulator \cite{Zdora2018}. UMPA is a flexible processing method that allows using different numbers of phase steps $N$ and enables to find a good trade-off between resolution and phase sensitivity by adjusting the analysis window size $w$ \cite{Zdora2017a}. A statistical analysis of the underlying principle yields the following expression for the variance of the differential phase image \cite{Zhou2016}:

\begin{equation}
    \sigma_x^2 = \frac{2\sigma_{\mathrm{ph}}^2}{\sum \left|\frac{\partial I}{\partial x}\right|^2}  = \frac{2\sigma_{\mathrm{ph}}^2}{N \Big \langle\left|\frac{\partial I}{\partial x}\right|^2 \Big \rangle}, 
\label{eq:Var_vx}
\end{equation}

where $\sigma_{\mathrm{ph}}^2$ is the variance of the photon noise, $N$ is the number of pixels in a window (in case of single-shot speckle tracking like in \cite{Zhou2016}) or the number of phase steps. $\big \langle... \big \rangle$ denotes the mean in a window or a pixel's phase stepping data set. Hence, an ideal speckle pattern should have strong intensity gradients i.e. small speckle sizes and high contrast. In the case of a periodic modulator, this is achieved at a high visibility-to-period ratio. Figure \ref{fig:FigSGradient} (a) and (b) show a comparison of partial derivatives in $x-$ and $y-$direction for the P1000 and the TAI 10-15 modulators (same data set as the visibility comparison in Figure 3 of the main article). The TAI modulation has both a stronger spatial gradient density as well as greater absolute gradient values. To compare the discussed modulators, we define a quantity $m = \big \langle \left|\frac{\partial I}{\partial x}\right|^2 \big \rangle^{-1}$ which scales linearly with $\sigma_x^2$ and can be calculated pixel-wise for any given intensity modulation and stepping scheme. Figure \ref{fig:FigSGradient} (c) shows histograms of $m$ values calculated for two TAIs and the P1000 diffuser from a ROI of 500 $\times$ 500 pixels. Both TAIs reach significantly lower $m$ values with a much narrower distribution compared to P1000. The mean of the P1000 (\num{1.49E-4}) is more than 6 times higher than that of TAI 10-15 (\num{0.246E-4}). Since $\sigma_{\mathrm{ph}}^2 \propto I$ \cite{Zhou2016} using the TAI instead of P1000 allows to reduce the dose and exposure time by a factor of 6 to reach comparable $\sigma_x$.  
Furthermore, Figure \ref{fig:FigSGradient} (c) shows that performing much more steps (e.g. 40) with the P1000 diffuser will reduce $m$ slightly and decrease the spread, however, will not reach comparable average gradients as the TAIs. Despite comparable visibility values (see Figure 3(c) in the main article) the TAI with 13 steps is still $5 \times$ more dose efficient than the P1000 diffuser with 40 steps ($m=\num{1.24E-4}$). In this regard, visibility values calculated by equation \ref{eq:visminmax} give a first, useful estimation of the modulator performance, but do not consider the spatial intensity gradient distribution, which is the key factor for a good phase sensitivity.

As evident from equation \ref{eq:Alpha} it is also possible to increase the propagation distance $d$ to achieve a higher phase sensitivity. This can be easily done with the proposed TAIs by going to a higher fractional Talbot distance (e.g. to 3/4$d_T$ for an increase by a factor of 4). In fact, different grating periods, duty cycles, and phase shifts provide even more flexibility to achieve optimal performance at a certain propagation distance and energy. However, too long propagation distances prove to be impractical for many samples as they result in far-field diffraction effects such as edge enhancement and deteriorate the modulation complicating phase retrieval. For some applications, however, where smooth phase gradients are of interest (e.g. wavefront sensing, optics characterization, etc.) going to higher fractional Talbot distances is a convenient solution. It should be also noted that with increasing propagation distance the visibility drops due to limited source coherence. In this regard, the TAIs also offer an advantage for laboratory-based systems as they reach high visibilities at short propagation distances.

A disadvantage of periodic modulators is an effect referred to as "phase wrapping" similar to grating-based imaging (GBI). Using a modulator of e.g. 10 pixels period and fitting a displacement of e.g. 4 pixels between the sample and reference scan it is not possible to distinguish from the neighboring spot which could have been shifted 6 pixels in the opposite direction. Hence, the dynamic range of the phase image is limited by the period. If a sample contains a high fraction of pixels affected by phase wrapping the range of sensitivity can be adapted by modifying $d$, the beam energy, or choosing a modulator with a bigger period. Furthermore, the maximum distance of the matched windows from sample and reference scan can be limited, and affected pixels can be discarded from the CT reconstruction. The same applies to sample regions where the modulation pattern is strongly deteriorated and cannot be matched reasonably to the reference pattern. This typically happens at strongly absorbing and scattering structures or at interfaces between air and soft tissue (e.g. sample vessel edges, air bubbles). 

The angular sensitivity $(\sigma_x, \sigma_y)$ reached in this work by 1D stepping was calculated from different scans performed with 13, 15, 17 and 25 steps and processed with window sizes of 3 and 5 pixels. For that the standard deviation (STD) of the signal in the differential phase images in a $ 150 \times 150$ pixel sample-free ROI was extracted and the sensitivity calculated by: 

\begin{equation}
\begin{pmatrix} \sigma_x \\ \sigma_y \end{pmatrix} = \frac{p_{\mathrm{eff}}}{d} \begin{pmatrix} \mathrm{STD(ROI_{dx}}) \\ \mathrm{STD(ROI_{dy}}) \end{pmatrix}. 
\label{eq:Sigma}
\end{equation}
Here, $p_{\mathrm{eff}}$ is the effective pixel size. All sensitivities are tabulated in Table \ref{tab:Sens}. The averaged visibility values were calculated pixel-wise from the respective flat-field steppings based on equation \ref{eq:visminmax}.

\begin{table}[htbp]
\centering
\caption{\bf Different image acquisition modes with respective visibilities and sensitivities.}
\begin{tabular}{ccccc}
\hline
Scan & N & Window [px] & Visibility & Sensitivity ($\sigma_x / \sigma_y$)[nrad] \\ 
\hline
Silica Particles  & 13 & 5  & 0.33 & 198 / 179   \\
Silica Particles  & 15 & 5  & 0.34 & 179 / 176   \\
Silica Particles  & 17 & 3  & 0.35 & 331 / 349   \\
Silica Particles  & 17 & 5  & 0.35 & 175 / 186   \\
Silica Particles  & 25 & 3  & 0.36 & 293 / 296   \\
Silica Particles  & 25 & 5  & 0.36 & 153 / 156   \\
\hline
Murine artery CT  & 15 & 3  & 0.42  & 199 / 217 \\
Murine artery CT  & 15 & 5  & 0.42  & 95 / 107  \\
\end{tabular}
  \label{tab:Sens}
\end{table}

\section{Estimation of resolution}
\begin{figure}[htbp]
\centering
\includegraphics[width=\linewidth]{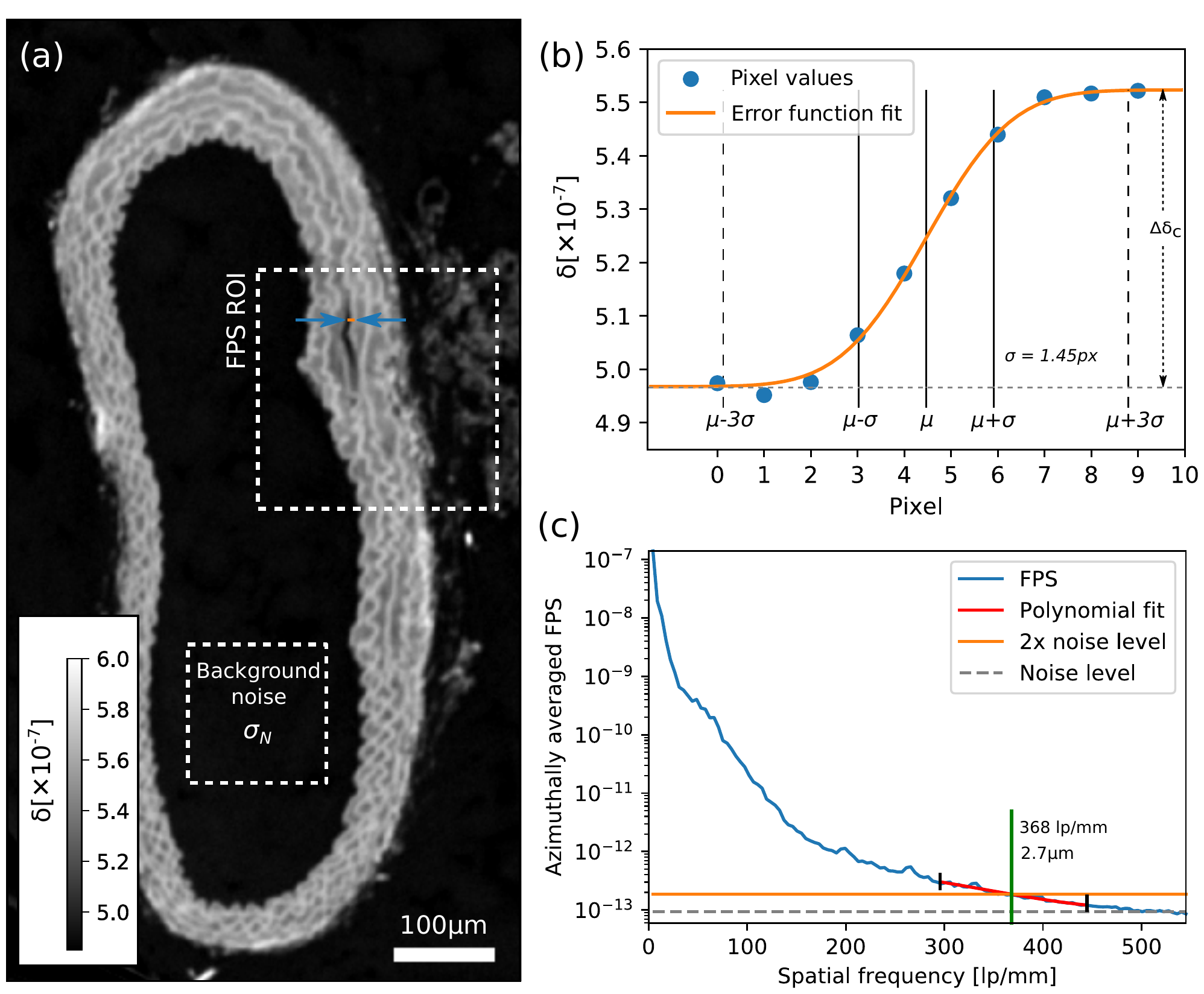}
\caption{(a) CT slice of the murine artery showing the ROI used for the FPS analysis, the position of an exemplary edge, and the ROI used for background noise determination. (b) The exemplary edge profile from the respective line plot in (a) fitted by the Gauss error function with pixel values in the range of at least $\pm 3\sigma$ around the mean $\mu$. (c) Plot of the azimuthally averaged FPS from the respective ROI in (a) showing the noise background level as well as the intersection of the FPS with the doubled noise level.}  
\label{fig:FigS4}
\end{figure}
The resolution of the murine artery CT scan was estimated with two methods. First, we evaluated sharp edges in the CT slices as exemplarily shown in Figure \ref{fig:FigS4}(a) by the orange line and depicted by blue arrows. Here the contrast between the paraffin matrix in which the artery was embedded and the vascular wall or the lamellar structures of the tunica media is of interest. For quantitative analysis, we implemented an automated edge detection algorithm. First, pixels with steep gradients of $\delta$ were detected by taking the spatial derivative of the image in the x-direction and applying a threshold filter. Then respective pixel values of the edges were automatically extracted and fitted by a Gauss error function as illustrated in Figure \ref{fig:FigS4}(b). Following criteria had to be present to select a clear edge: (1) The contrast $\Delta\delta_c$ of the edge had to be at least \num{0.3E-7}, which is more than $20\times$ higher than the noise level of the background paraffin wax matrix ($\sigma_N =\num{0.015E-7}$). (2) The edge had to consist of a number of pixels covering $\pm 3\sigma$ around the mean $\mu$ and (3) the residuum (measurement-fit) of every pixel value had to be less than 5\% of the contrast $\Delta\delta_c$ to exclude badly shaped edges e.g. with pixel outliers. The algorithm was performed in 20 equidistant slices all over the CT volume of interest and from every slice 10 edges with the lowest $\sigma$ values were extracted. From those 200 valid edges an average of $\sigma = 1.43 \pm 0.088$ pixels was found resulting in an average FWHM of $3.06 \pm 0.19$ \si{\micro\meter} which we consider as the spatial resolution of the scan.
The second method based on the Fourier power spectrum (FPS) was proposed in \cite{Modregger2007}. A ROI (see Figure \ref{fig:FigS4}(a)) covering some structure of the vessel as well as the paraffin wax matrix was chosen for further analysis. The FPS was calculated according to \cite{Modregger2007}, azimuthally averaged, and plotted in Figure \ref{fig:FigS4}(c). The noise level depicted by the dashed line was determined from the average of the last 10 frequency bins and the doubled noise level is plotted as the orange line. To determine its intersection with the FPS the latter was fitted with a 2nd order polynomial around the region of interest indicated by the red line (fit) with the respective black crossbars. The intersection depicted by the green line lies at 368 lp/mm and corresponds to a feature size of \SI{2.7}{\micro\meter}. 

{\small \bibliography{references_sup}} 